\documentclass[reprint,bibnotes,amsmath,amssymb,showpacs,floatfix,superscriptaddress,bibliography]{revtex4-2}

\usepackage{amsmath,empheq}
\usepackage{amsthm}
\usepackage{amsfonts}
\usepackage{amssymb}
\usepackage{amsxtra}
\usepackage{mathtools}
\usepackage{xcolor}
\usepackage{graphicx}
\usepackage{subfigure}
\usepackage{dcolumn}
\usepackage{mathrsfs}
\usepackage{tikz}
\usepackage{float}
\usepackage{bm}
\usepackage[breaklinks=true,colorlinks,citecolor=purple,linkcolor=purple,urlcolor=purple]{hyperref}

\DeclareMathAlphabet{\bi}{OML}{cmm}{b}{it}
\def\be{\begin{equation}}
	\def\ee{\end{equation}}
\def\bearr{\begin{eqnarray}}
	\def\eearr{\end{eqnarray}}

\begin{document}
	\title{Optical pumping controls anisotropic response in semi-Dirac system}
	\author{Bristi Ghosh}
	\email{a21ph09025@iitbbs.ac.in}
	\author{Malay Bandopadhyay}
	\email{malay@iitbbs.ac.in}
	\affiliation{School of Basic Sciences, Indian Institute of Technology Bhubaneswar, Argul, Jatni, Khurda, Odisha 752050, India}
	\author{Ashutosh Singh}
	\email{asingh.n19@gmail.com}
	\affiliation{Department of Physics and Astronomy, Texas A\&M University, College Station, TX, 77843 USA}
	\date{\today}
	
	\begin{abstract}
		Low-energy Fermions in semi-Dirac systems depict linear momentum dispersion along one direction while having the features of parabolic dispersion in the other direction. Equilibrium optical responses of such highly anisotropic dispersion are manifested in direction-dependent optical conductivity tensor. Going beyond the equilibrium framework, here we probe the effects of optical pumping-led non-equilibrium carrier distribution on this system's transmission and polarization rotation. Within the equation of motion approach for a two-band density matrix, we obtain a quasi-steady state solution for a continuous wave (CW) illumination, in which the population of the two bands is characterized by non-thermal occupancy factors with a strong dependence on the amplitude, polarization, and the frequency of the pump field. We demonstrate that tuning the pump field parameters significantly modifies the optical conductivity tensor for the probe field, which can have important practical consequences such as selective transmission and tunable hyperbolicity.
	\end{abstract}
	
	%
	\maketitle
	%
\section{Introduction}
Questions concerning band properties and interaction effects in condensed matter systems have been discussed within the linear response framework for the most part\cite{Mahan1990}, even though nonlinear optics offers a broader range of possibilities \cite{Boyd2008, Gunter}. With technological advancement, it has become possible to engineer high-power laser sources and detectors that can facilitate the observation of nonlinear response to a very high degree of accuracy. Accordingly, using the nonlinear optical response as a tool for probing and controlling quantum materials has become a popular subject recently \cite{Basov2017, https://doi.org/10.1002/adma.201705963, annurev:/content/journals/10.1146/annurev-conmatphys-031218-013712}. A major part of this progress was envisioned after the discovery of graphene\cite{Geim2007}. Not surprisingly, several studies were followed that explored the nonlinear optical properties of graphene\cite{PhysRevLett.105.097401, Mikhailov_2007, PhysRevLett.103.246802, PhysRevLett.112.055501}. It was pointed out that the relativistic nature of the quasiparticle dispersion allows for efficient nonlinear pathways giving rise to THz plasmons\cite{PhysRevLett.112.055501} and generation of entangled photons\cite{PhysRevLett.110.077404, PhysRevLett.108.255503}. Among other major developments were the topological control of the nonlinear response due to strong optical fields\cite{PhysRevB.79.081406, annurev:/content/journals/10.1146/annurev-conmatphys-031218-013423, doi:10.1126/sciadv.1501524}. Very recently it was demonstrated that chiral currents in the quantum Hall regime can be modified via nonlinear optical rectification mechanism\cite{singh2024coherent}.\\
\begin{figure}[!]
\includegraphics[width =\linewidth]{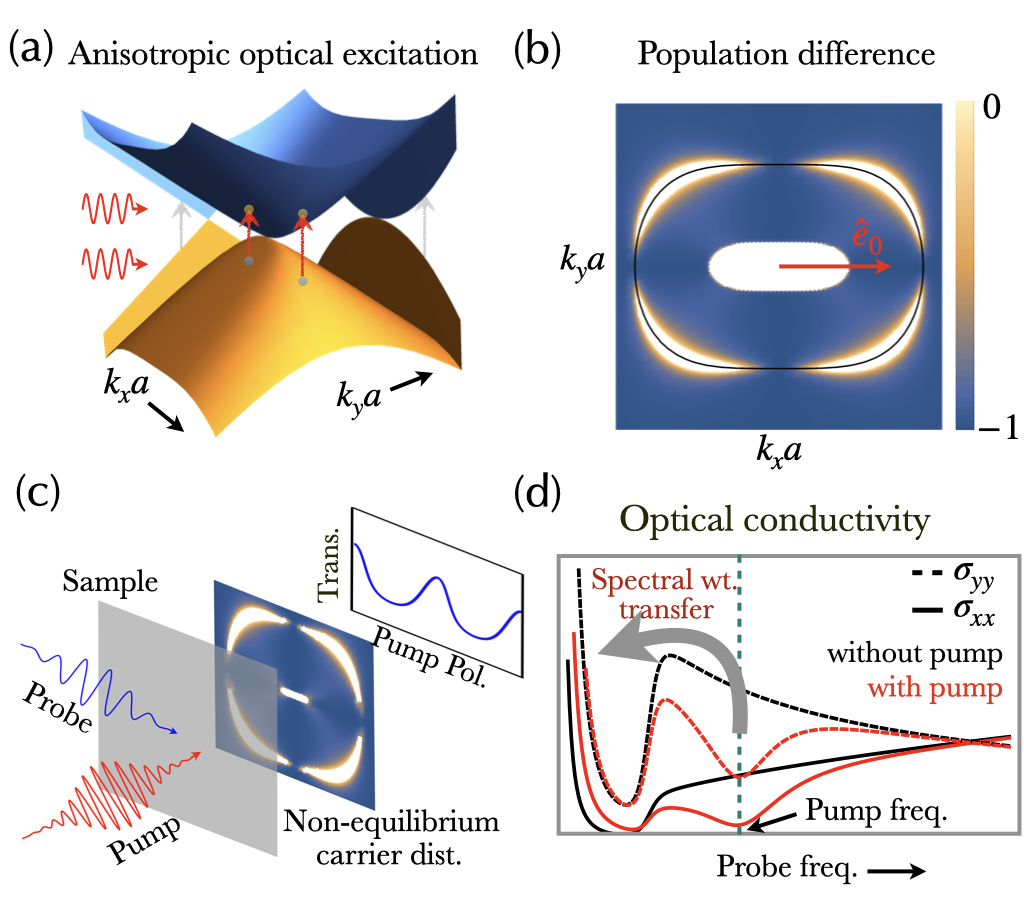}
\caption{(a). Optical excitations for anisotropic energy dispersion for the Hamiltonian in Eq.~\eqref{Ham} as a function of $k_xa$ and $k_ya$. Projection on the energy-momentum plane shows the Dirac ($k_x=0$) and the parabolic ($k_y=0$) dispersion. (b) In the presence of a strong pump field, the carrier distribution function changes from a Fermi-Dirac distribution to a non-equilibrium distribution function which depends on the field parameters (such as polarization direction $\hat {\bf e}_0$). The white region in the center is due to Pauli blocking which prohibits optical excitation below the doping level. The black curve represents the momentum points at which the transition frequency equals the pump frequency. (c) The transmission profile for the probe field strongly depends on the pump field parameters. (d) The anisotropic dispersion leads to differences in the magnitude as well as frequency dependence of longitudinal optical conductivities, which can be modified utilizing optical pumping.}\label{fig_dispersion}
\end{figure}
\indent
The growth of activities focused on systems with relativistic fermions, commonly known as Dirac and Weyl materials, stems from recent developments in material research. In a rather peculiar scenario, there exists a class of materials hosting quasiparticles with energy dispersion illustrating semi-Dirac features\cite{PhysRevLett.102.166803, PhysRevB.86.075124, doi:10.1126/science.aaa6486, Baik2015, shao2023semidiracfermionstopologicalmetal}, that is, massive as well as mass-less fermions coexist in such systems. This can happen if the Dirac points move in the Brillouin zone and eventually merge as a result of changing hopping parameters \cite {PhysRevB.80.153412, Montambaux2009, Tarruell2012, PhysRevB.93.125113}. Such merging and gap opening can also happen via the application of strong irradiation giving rise to Chern insulating states\cite{PhysRevB.88.245422, PhysRevB.92.161115}. The effect of anisotropic band structure and the transition from semi-metallic to insulating phase has been investigated for plasmon dispersion\cite{PhysRevB.93.085145}, linear optical conductivity\cite{Ziegler_2017, Sanderson_2018,PhysRevB.99.115406, PhysRevB.99.075415, PhysRevB.106.115143,PhysRevB.100.035441}, tunable transmittance \cite{PhysRevB.97.235411, PhysRevB.105.115423}, and anisotropic absorption\cite{PhysRevB.105.115423}. Differences in the longitudinal conductivity in two directions in the semi-Dirac system can also lead to hyperbolic electromagnetic field propagation \cite{rudenko2024anisotropiceffectstwodimensionalmaterials}. It was demonstrated recently that hyperbolic polaritons can emerge in nodal ring semimetals \cite{PhysRevLett.131.096902}, due to a similar kind of anisotropy in the electron dispersion. Moreover, linear and nonlinear magneto-optical studies for semi-Dirac systems have also been carried out\cite{PhysRevB.104.235403, PhysRevB.105.205407, PhysRevB.108.035203}, due to its characteristically different Landau quantisation\cite{PhysRevB.80.153412, Montambaux2009} as compared to purely parabolic or relativistic fermionic dispersion. 
\\
\indent
Within the Kubo formulation, optical response as a function of photon frequency for finite doping is usually characterized by a Drude peak close to zero frequency and then attaining a finite value for a frequency larger or equal to twice the doping level (Fig.\ref{fig_dispersion}(d)). The setup for nonlinear response studies typically involves the perturbative solution of the equation of motion for density matrix \cite{Boyd2008}, with the diagonal elements of the density matrix denoting the population and the off-diagonal elements denoting inter-band coherence. In the coherent regime, the set of equations describing the carrier dynamics is called as the optical Bloch equation (OBE)\cite{PhysRevB.52.14636, Koch2004}. The interaction effects can be incorporated for simplicity by including damping terms in the OBE. Electron-photon interaction terms in the OBE for two-band systems are described by momentum-dependent intra- and inter-band Rabi frequencies, carrying within them the information of field parameters and transition matrix elements. The remaining two energy scales in the OBE for vertical optical transitions are photon energy and inter-band transition frequency. Concentrating on inter-band dynamics and eliminating intra-band Rabi terms from the OBE is crucial for obtaining the steady-state analytical solution for the density matrix elements in the rotating-wave approximation (RWA). Due to its analytical tractability, this approach has been employed for graphene in the context of nonlinear conductivities and plasmonics\cite{PhysRevLett.103.246802, PhysRevB.93.041413, PhysRevB.94.195438, PhysRevB.95.155421, PhysRevB.97.045402, PhysRevB.99.125419, PhysRevB.97.205420}.\\
\indent
In this work, we focus on the effect of strong continuous wave (CW) illumination on the anisotropic transmission and polarization rotation in two-band semi-Dirac Hamiltonian. Although several fascinating features exploring the anisotropic dispersion of the semi-Dirac systems have been studied, the areas that involve non-equilibrium carrier distribution guided tunable transmission and polarization rotation are relatively unexplored to the best of our knowledge. Unlike other anisotropic systems \cite{PhysRevB.97.205420}, semi-Dirac dispersion offers a qualitatively distinct momentum profile of the non-equilibrium carrier distribution generated by a pump field \cite{Dai_2019}.
In addition to the pump field, we have added a probe field to the system which carries the signature of such an anisotropic carrier distribution in its optical conductivity tensor. We further explore possible ways to experimentally realize this nonlinear anisotropic response through the transmission profile of the probe field. \\ 
\indent
With this preamble, the rest of the paper is organized as follows: In Sec. \ref{method}, we present the theoretical method discussing the two-band semi-Dirac Hamiltonian and the formulation for obtaining the steady-state solution of the OBE under CW illumination. In the first part in Sec. \ref{Op_conductivity} we calculate $xx$- and $yy$- components of the optical conductivity in the linear regime and provide the analytical results. In the later part, we present numerical results for the longitudinal as well as transverse nonlinear optical conductivities and obtain corresponding analytical results in the perturbative limit, keeping the leading order term that modifies linear conductivities. In Sec. \ref{sec:Transmission}, we discuss the effects of nonlinearity on the transmission corresponding to $s$- and $p$- polarized incidence. Details of the calculation and discussion on the steady-state solution are provided in the appendix part.
\section{Theoretical Method}\label{method}
\subsection{Electron dispersion}
\begin{figure}[!]
\includegraphics[width =0.9\linewidth]{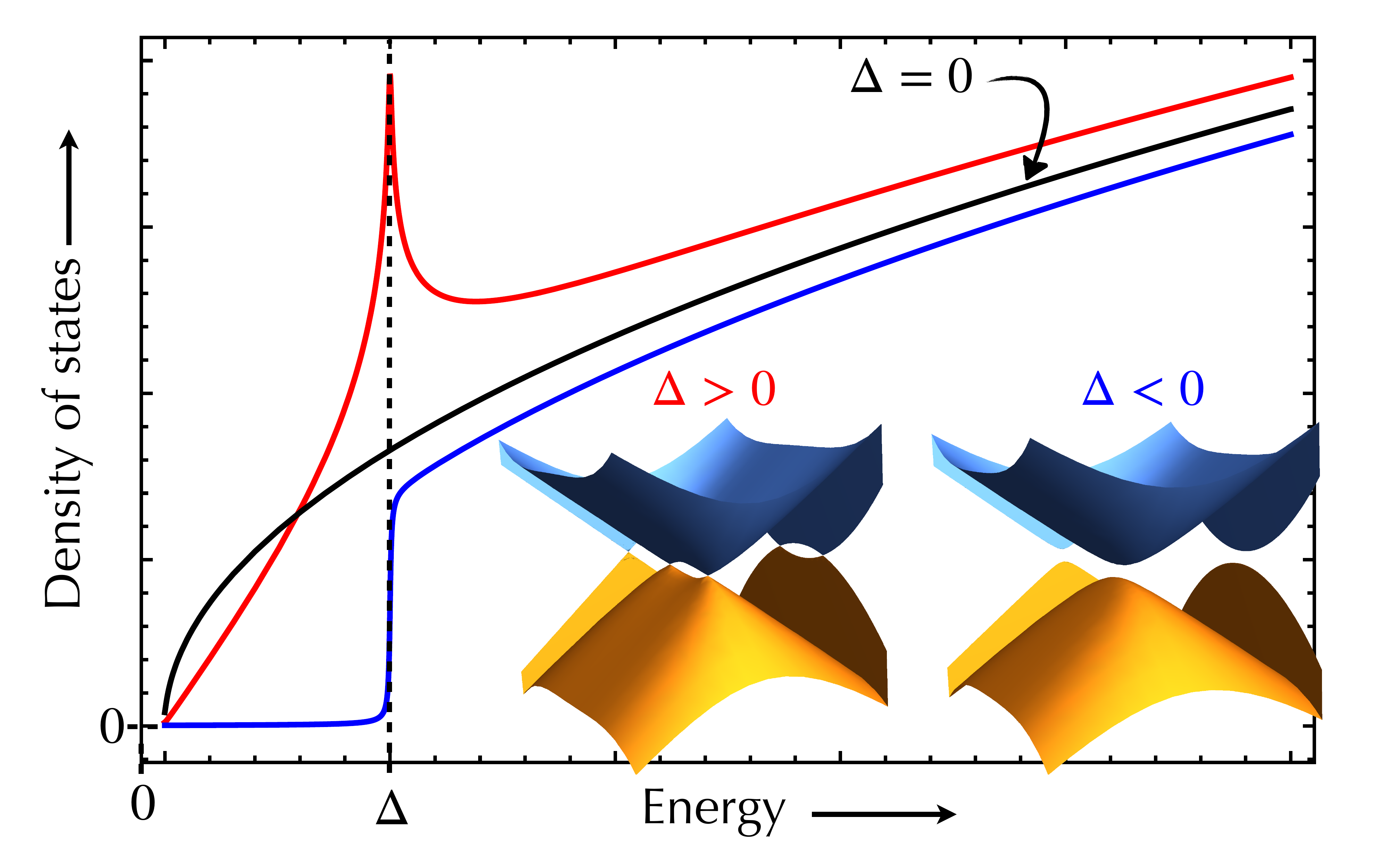}
\caption{The density of states (DOS) as a function of energy computed for the Hamiltonian in Eq.~\eqref{Ham}. Red, blue, and black curves represent the positive values, negative values, and zero of the gap parameter $\Delta$, respectively. As energy crosses $\Delta$, DOS changes abruptly for the red curve as it encounters the Van-Hove singularity. No such singularity exists when $\Delta$ is negative.}\label{DOS}
\end{figure}

In this work, we consider a low-energy Hamiltonian illustrating parabolic dependence on momentum in one direction whereas having linear dependence on momentum in the other direction. For keeping the discussion general, we can describe the Hamiltonian describing two different phases separated by a tuning parameter $\Delta$, given as,
%
%
\begin{align}\label{Ham}
\hat H_{0\bf k} = \left(tk_x^2a^2/2 -\Delta\right)\hat\sigma_x + \alpha tk_y a\hat\sigma_y~,
\end{align}
where $\hat\sigma_x$ and $\hat\sigma_y$ are the Pauli matrices, ${\bf k}=(k_x, k_y)$ is the momentum
vector, $a$ is the lattice parameter, 2$\Delta$ being the gap at momentum ${\bf k}=(0,0)$ and $t$ and $\alpha t$ are the hopping parameters in the $x$ and $y$ direction respectively. Here $\alpha (> 0)$ is introduced as the anisotropy parameter.
The energy eigenvalues for the conduction and the valance bands are given as,
\begin{align}
\varepsilon_{\lambda{\bf k}} = \frac{\lambda}{2} t \sqrt{(\tilde k_x^2 - 2\tilde\Delta)^2+4 \alpha ^2 \tilde k_y^2},
\end{align}
for $\lambda = c/v (=\pm)$ for conduction/valance band respectively. For simplicity, we define $\tilde k_x = k_x a$, $\tilde k_y = k_y a$ and $\tilde\Delta=\Delta/t$ for the dimensionless momentum and the gap parameter. We can define transition frequency corresponding to a vertical transition given as,
$\hbar\omega_{\bf k} = \varepsilon_{c{\bf k}}-\varepsilon_{v{\bf k}} = t \sqrt{(\tilde k_x^2 - 2\tilde\Delta)^2+4 \alpha ^2 \tilde k_y^2}$.
Next, the normalized eigenstate for the conduction and the valance bands are given as,
\begin{eqnarray}
|\psi_{\lambda\bf k}\rangle&=&\frac{1}{\sqrt{2}}\left(\lambda\frac{\sqrt{(\tilde k_x^2-2\tilde\Delta)^2+4 \alpha ^2 \tilde k_y^2}}{\left(\tilde k_x^2+2 i \alpha\tilde k_y-2\tilde\Delta  \right)},1\right)^{\rm T}~.
\end{eqnarray}
In Fig.~\eqref{fig_dispersion}-\eqref{DOS} we show the energies for the conduction and the valence band on the momentum plane for the two phases, in which the anisotropic nature of the low-energy quasiparticles are obvious. Two nodes that appear for non-vanishing $\Delta$, tend to merge as $\Delta\to 0$. Furthermore, the finite gap in the energy dispersion can appear for $\Delta\to -\Delta$. For these cases, the density of states as a function of energy is shown in Fig.~\eqref{DOS}.
\subsection{Optical excitation}
In dipole approximation, the total Hamiltonian in the presence of electromagnetic field can be written as,
\begin{align}
H = H_{0\bf k} + e{\bf E}\cdot{\hat{\bf r}}~,
\end{align}
with ${\hat{\bf r}}$ being the position operator, and the electric field $ {\bf E}= \hat {\bf e}_0 (E_0(\omega)e^{-i\omega t}/2 + c.c.) $, where the polarization vector $\hat {\bf e}_0 = (\cos{\theta},\sin{\theta}) $ with $\theta$ being the angle w.r.t. $\hat x$-axis. Off-diagonal elements of the position operator represented in the eigen-basis of $H_{0\bf k}$, that are important for inter-band transitions, can be recasted in their corresponding optical matrix elements with the general definition thereof ${\mathcal M}^{\lambda\lambda^{\prime}}_{\bf k} = e \hbar^{-1}\langle \psi_{\lambda\bf k}| \nabla_{\bf k}H_{0\bf k}|\psi_{\lambda^{\prime}\bf k}\rangle$ \cite{PhysRevB.95.155421, PhysRevB.97.045402, PhysRevB.99.125419}. Thus the elements of the optical matrix are,
\begin{eqnarray}
\frac{{\mathcal M}^{cc}_{\bf k}}{eat/\hbar} &=&\frac{\left(\tilde k_x (\tilde k_x^2-2\tilde\Delta),2 \alpha ^2 \tilde k_y\right)}{\sqrt{(\tilde k_x^2-2\tilde\Delta)^2+4 \alpha ^2 \tilde k_y^2}}~,\\
\frac{{\mathcal M}^{vc}_{\bf k}}{eat/\hbar} &=&  \frac{\left(-2 i \alpha  \tilde k_x \tilde k_y,i \alpha  (\tilde k_x^2-2\tilde\Delta)\right)}{\sqrt{(\tilde k_x^2-2\tilde\Delta)^2+4 \alpha ^2 \tilde k_y^2}}~,\\
{\mathcal M}^{vv}_{\bf k} &=& -{\mathcal M}^{cc}_{\bf k} ~,\\
{\mathcal M}^{cv}_{\bf k} &=& \left({\mathcal M}^{vc}_{\bf k}\right)^*~,
\end{eqnarray}
To describe the dynamics of the charge carriers, we use the equation of motion for the density matrix,
\begin{align}
i\hbar\frac{\partial \hat\rho_{\bf k}(t)}{\partial t} = \left[H, \hat\rho_{\bf k}(t)\right],
\end{align}
where, $\hat\rho_{\bf k}(t) = \sum_{\lambda,\lambda^{\prime}}\rho^{\lambda\lambda^{\prime}}_{\bf k}(t)|\psi_{\lambda\bf k}\rangle\langle \psi_{\lambda^{\prime}\bf k}|$. Denoting $p_{\bf k} = \rho^{cv}_{\bf k}$ as the interband coherence and $n_{\bf k} = \rho^{cc}_{\bf k} - \rho^{vv}_{\bf k}$ as the population difference between the conduction and the valance band, we obtain the following set of
Optical Bloch equation,
\begin{eqnarray}\label{OBE1}
\frac{\partial n_{\bf k}}{\partial t} &=& 4{\rm Im}\left[\Omega^{cv}_{\bf k}p_{\bf k}\right] - \gamma_{1}\left(n_{\bf k}-n^{\rm eq}_{\bf k}\right)~,\\
\frac{\partial p_{\bf k}}{\partial t} &=& i\omega_{\bf k}p_{\bf k} - i\Omega^{vc}_{\bf k}n_{\bf k} - \gamma_2p_{\bf k}\label{OBE2}~,
\end{eqnarray}
where we have introduced $\gamma_{1}$ and $\gamma_{2}$ as the phenomenological damping terms to incorporate interaction effects, $\Omega^{\lambda\lambda^{\prime}}_{\bf k} = e{\bf E}\cdot{\bf r}^{\lambda\lambda^{\prime}}/\hbar$ is the Rabi frequency, and $n^{\rm eq}_{\bf k} = f_{c\bf k} - f_{v\bf k}$ with $f_{\lambda\bf k} = \left(1 + e^{(\varepsilon_{\lambda\bf k}-\mu)/(k_BT)}\right)^{-1}$ at temperature $T$ and chemical potential $\mu$. Analogous to the Drude contribution, which remains localized in the low-frequency regime for low temperature and weak impurities case, intra-band Rabi terms in Eqs.~\eqref{OBE1}-\eqref{OBE2} do not play significant role at high frequencies and we have dropped them in order to proceed analytically.
For an ansatz $p_{\bf k} = p_{1\bf k}e^{i\omega t} + p_{2\bf k}e^{-i\omega t}$, the steady-state population inversion ($\partial_t n_{\bf k} = 0$) is given as \cite{PhysRevB.95.155421},
%
%
%
 %
 \begin{equation}\label{n_k}
 \frac{n_{\bf k}}{n^{\rm eq}_{\bf k}}=  \left(1+\eta^2\frac{\omega^2}{\omega^2_{\bf k}}\frac{|{\mathcal M}^{vc}_{\bf k}\cdot \hat {\bf e}_0|^2 }{(eat/\hbar)^2} \sum_{s=\pm} {\mathcal W}_s(\omega_{\bf k},\omega)\right)^{-1}~.
 \end{equation}
where, ${\mathcal W}_s(\omega_{\bf k},\omega) = \gamma^2_2/(\gamma^2_2 + (\omega_{\bf k}-s\omega)^2)$, and $\eta=\frac{e a t E_0}{\hbar^2 \omega \sqrt{\gamma_1 \gamma_2}}$.  We can see that ${\mathcal W}_s(\omega_{\bf k},\omega)/\gamma_2$ has a Lorentzian profile, which is centered at $\omega_{\bf k} = s\omega$. Since $\omega_{\bf k}$ is always positive, ${\mathcal W}_+(\omega_{\bf k},\omega)$ is much larger than ${\mathcal W}_-(\omega_{\bf k},\omega)$ at high enough frequencies. The phenomenological damping terms contain information about various types of interaction and impurity effects and therefore depend on momentum, frequency, and temperature \cite{Semnani2019,10.1088/1402-4896/ad7ba1}. For simplicity and analytical tractability, we assume $\gamma_1$ and $\gamma_2$ to be constant. Furthermore, the non-linearity parameter $\eta$ is a dimensionless quantity that is given by the ratio of the maximum value of the inter-band Rabi frequency and the effective decay rate $\sqrt{\gamma_1\gamma_2}$. Defined this way, $\gamma_1$ disappears from the expression for the carrier population.  On the other hand, $\gamma_2$ not only appears in $\eta$ but also determines the transition line-width.
%
%
%
%
\begin{figure}[t!]
\includegraphics[width =\linewidth]{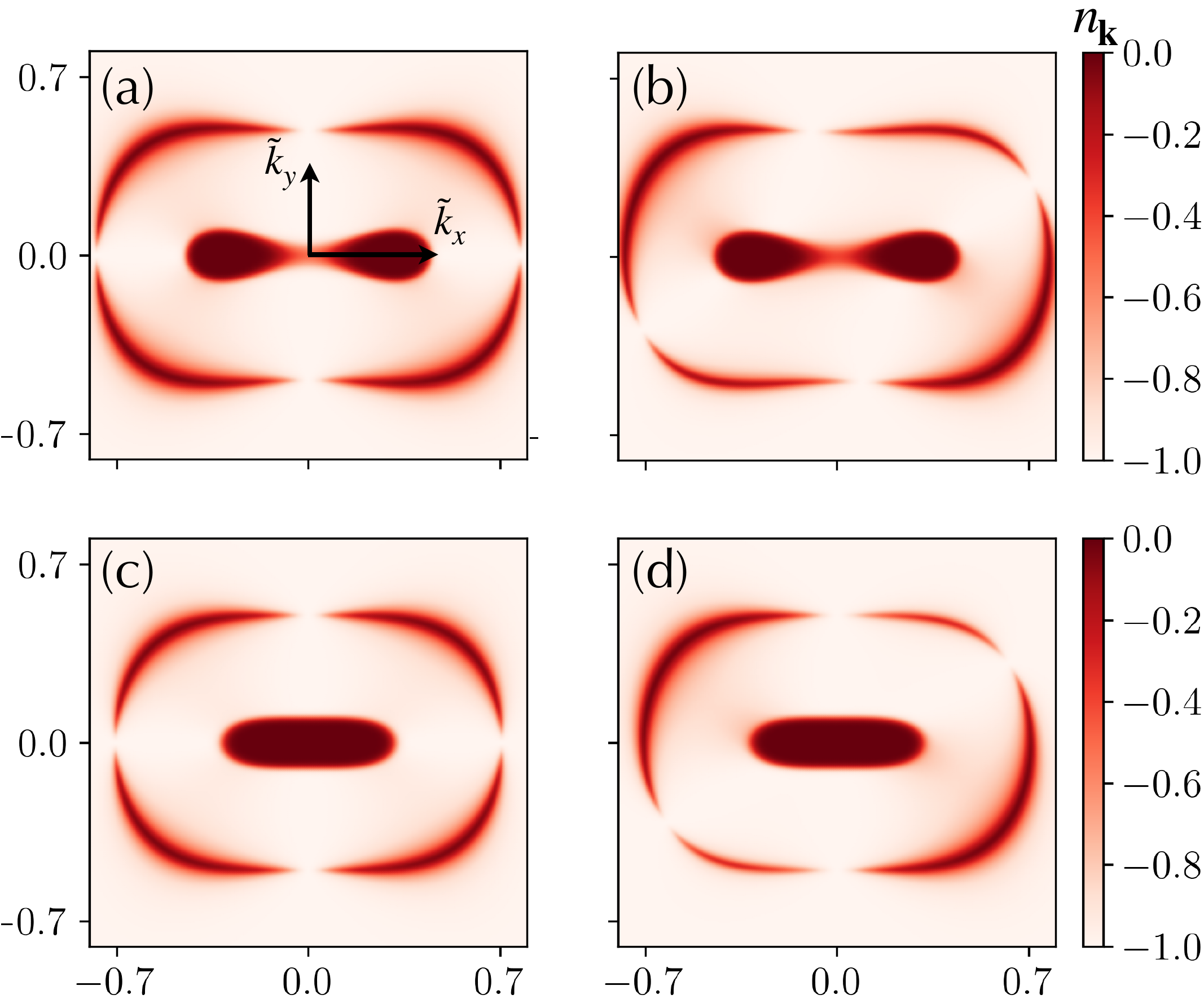}
\caption{Momentum dependence of $n_{\bf k} $ for $\alpha = 1/2$ in the top panel for $\Delta = 0.05t$ and (a) $\theta = 0$, (b) $\theta = \pi/4$. In the bottom panel we have $\Delta = 0$, and (c) $\theta = 0$, (d) $\theta = \pi/4$. $n_{\bf k}\to -1$ corresponds to the no optical pumping case and $n_{\bf k}\to 0$ refers to the population saturation. The dark region in the center is due to Pauli blocking. The anisotropy of the dispersion affects the nonlinear optical regime significantly. Here $\omega = 0.5t/\hbar, \hbar\gamma_2 = 0.01 t, \mu/t=0.05$, $\eta = 10$ and $t/(k_BT) = 300$.} \label{fig:n_k}
\end{figure}

In Fig.~\ref{fig:n_k} we have shown the anisotropic nature of the carrier population difference as a function of two-dimensional momentum. For the positive finite value of the gap parameter, the conduction and the valence band tend to touch as $(\tilde k_x, \tilde k_y)\to (\pm \sqrt{2\Delta/t},0)$ whereas the saddle point in the dispersion is located at the origin.The dark region in the vicinity of these Dirac points is due to Pauli blocking, since all states are occupied below the doping level. Under the action of the pump field, the quasi-steady state regime supports the net transfer of carriers from the valance band to the conduction band. Due to the specific nature of the pseudo-spin-momentum locking, the momentum profile of the population difference is composed of the dips and peaks within the same anisotropic energy contour (Fig.~\ref{fig:n_k}(a)). Furthermore, at finite values of the pump polarization angle, the momentum profile of the population becomes asymmetric for $\tilde k_{x/y}\to -\tilde k_{x/y}$, as shown in Fig.~\ref{fig:n_k}(b). Merging of these two Dirac points takes place for zero value of the gap parameter, and now the conduction and the valance band tend to touch as $(k_x, k_y)\to (0,0)$. In this case, the population momentum profile closely follows the previous situation except near the origin as shown in Fig.~\ref{fig:n_k}(c)-(d).

\section{Optical Response}\label{Op_conductivity}
%
%
In this section we first discuss the optical conductivity in the context of our non-equilibrium population, which is different from the thermal distribution. The optical conductivity for a probe field ${\bf E}_{\rm pr}e^{-i\omega_{\rm pr}t}+c.c.$, in the presence of the pump field can be defined as,
\begin{align}
\sigma_{ij}(\omega_{\rm pr};\omega) = g_s\int \frac{d{\bf k}}{(2\pi)^2}\frac{J^i_{\bf k}(\omega_{\rm pr};\omega)}{E^j_{\rm pr}(\omega_{\rm pr})}~.
\end{align}
where, $E^j_{\rm pr}(\omega_{\rm pr})$ is the $j$-th component ($j = x,y$) of the electric field vector, ${\bf E}_{\rm pr}$, $J^i_{\bf k}$ is the $i$-th component ($i = x,y$) of the current density,
\begin{eqnarray}\label{current}\nonumber
{\bf J}_{\bf k}(\omega_{\rm pr};\omega) &=& \frac{i n_{\bf k}}{\hbar\omega_{\bf k}}\left[\frac{({\bf E}_{\rm pr}\cdot {\mathcal M}^{cv}_{\bf k}){\mathcal M}^{vc}_{\bf k}}{\omega_{\bf k} -\omega_{\rm pr} - i\gamma} - \frac{({\bf E}_{\rm pr}\cdot {\mathcal M}^{vc}_{\bf k}){\mathcal M}^{cv}_{\bf k}}{\omega_{\bf k} + \omega_{\rm pr}+ i\gamma} \right]\\
&-&\frac{2i}{\hbar}\frac{\partial n_{\bf k}}{\partial\omega_{\bf k}}\frac{({\bf E}_{\rm pr}\cdot {\mathcal M}^{cc}_{\bf k}){\mathcal M}^{cc}_{\bf k}}{ \omega_{\rm pr} + i\gamma}~,
\end{eqnarray}
and $g_s$ is the spin degeneracy factor. Here, the first and second terms correspond to the inter-band and intra-band optical transition, respectively. As was shown in Ref.\cite{PhysRevLett.131.096902}, the anisotropy in the dispersion leads to huge differences in the diagonal components of intra-band optical conductivities, resulting in
hyperbolic dispersion of polaritons. In the present scenario, we can control the hyperbolic profile of the polaritons using the parameters of the pump field.
However, we are mainly interested in the inter-band optical conductivities that dominate at high frequencies. Keeping only the dominant term, the optical conductivity is given as,
\begin{align}\label{conductivity}
\frac{\sigma_{ij}(\omega_{\rm pr};\omega) }{g_s/(4\pi^2\hbar)}= i\int \frac{d{\bf k}}{\omega_{\bf k}}\left[\frac{n_{\bf k}({\mathcal M}^{vc}_{{\bf k}i}{\mathcal M}^{cv}_{{\bf k}j})}{\omega_{\bf k} -\omega_{\rm pr} - i\gamma} - \frac{n_{\bf k}({\mathcal M}^{vc}_{{\bf k}j}{\mathcal M}^{cv}_{{\bf k}i})}{\omega_{\bf k} + \omega_{\rm pr}+ i\gamma} \right]~.
\end{align}
\subsection{Linear regime ($\eta\to 0$, $n_{\bf k}\to n^{\rm eq}_{\bf k}$)}
To start with, we consider a situation where the pump field is absent. Within this condition, Kubo formalism for linear response is applicable. The analytical results were obtained earlier in the limiting cases of small and large frequency in Ref. \cite{PhysRevB.100.035441}. Here we provide the full frequency dependence of the conductivities in the linear response regime which is obtained from our general expression for nonlinear response in the $\eta\to 0$ and $\gamma\ll\omega_{\rm pr}$ such that $n_{\bf k}\to n^{\rm eq}_{\bf k}$. The real part of the conductivity is,
\begin{align}\label{conductivity_Kubo}
{\rm Re}[\sigma^{\eta\to 0}_{ij}(\omega_{\rm pr})] = -g_s\int \frac{d{\bf k}}{(2\pi)^2}\frac{ n^{\rm eq}_{\bf k}}{\hbar\omega_{\bf k}}{\mathcal M}^{vc}_{{\bf k}i}{\mathcal M}^{cv}_{{\bf k}j}\pi\delta(\omega_{\bf k} -\omega_{\rm pr}) ~.
\end{align}
Due to the anisotropic band structure, it is a difficult task to compute the above integral. To simplify this, we can convert,
\begin{align}
\delta(\omega_{\bf k}-\omega)\to \sum_{\ell=\pm}\frac{\delta(k_y-k^{\ell}_{\omega})}{|\partial_{k_y}\omega_{\bf k}|_{k_y\to k^{\ell}_{\omega}}},
\end{align}
where, $k^{\ell}_{\omega} = \ell\sqrt{\hbar^2\omega ^2-\left(a^2k_x^2 t-2 \Delta \right)^2}/(2 \alpha a t)$. This makes the $k_y$ integral trivial. Furthermore, $k_x$ integration is bounded due to the condition $:\hbar^2\omega ^2-\left(a^2k_x^2 t-2 \Delta \right)^2 \geq 0$. For $\hbar\omega < 2\Delta$, we have $\sqrt{2\Delta-\hbar\omega} < a\sqrt{t}|k_x| < \sqrt{2\Delta+\hbar\omega}$, whereas for $\hbar\omega \geq 2\Delta$,  $0< a\sqrt{t}|k_x| < \sqrt{2\Delta+\hbar\omega}$. After performing the $k_x$ integral, the expression for the real parts of the $xx$- and $yy$- components of conductivities are given as, 
\begin{figure}[h]
\includegraphics[width =0.9\linewidth]{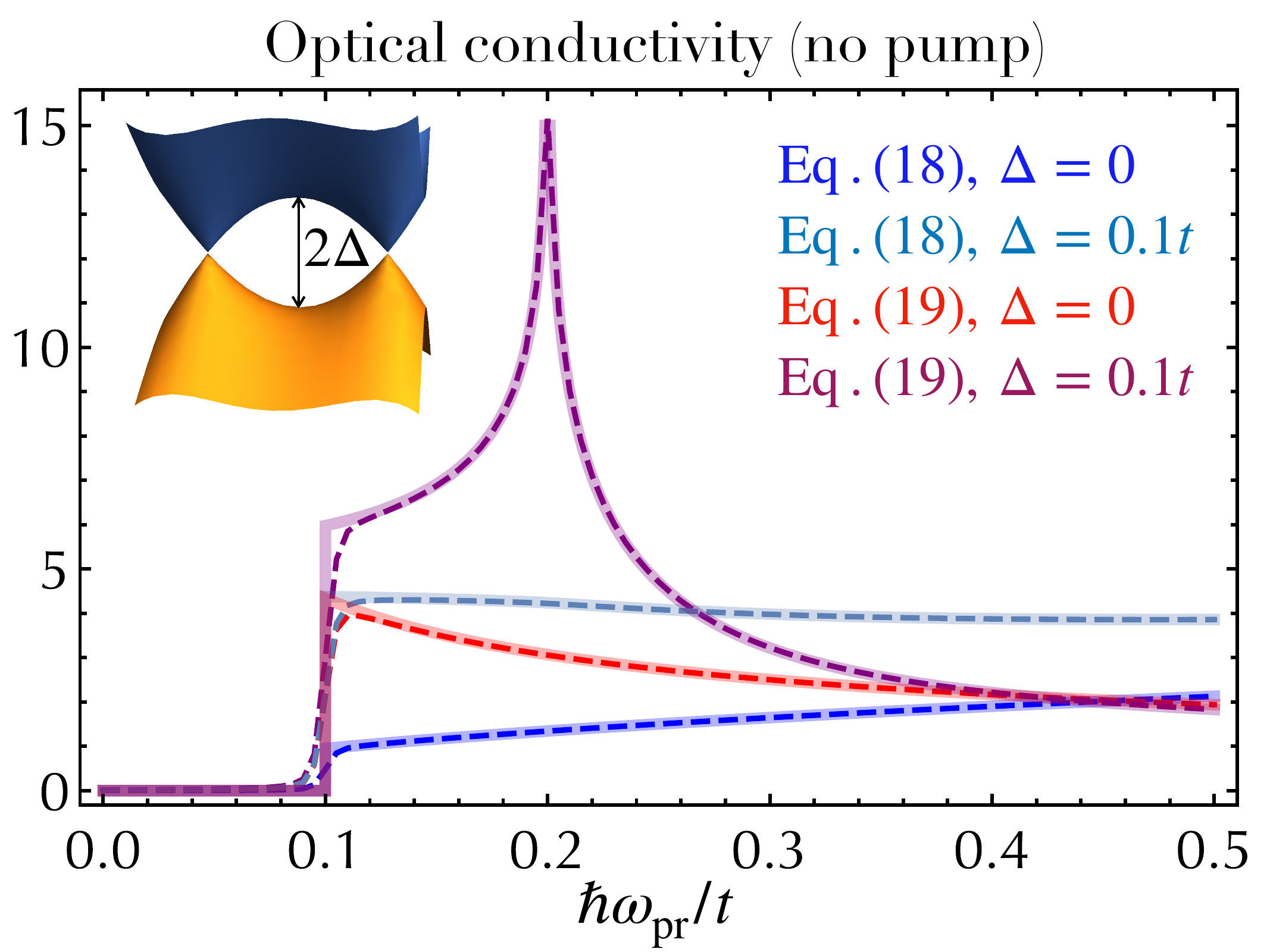}
\caption{Numerical (Eq.~\ref{conductivity_Kubo}, dashed lines) and analytical results (solid lines) for the real parts of normalized optical conductivity ($\sigma^{\eta\to 0}_{ij}/\sigma_0$) w.r.t. normalized probe photon energy in the absence of the pump field. Here $\sigma_0 = e^2/(\pi h)$,  and for the numerics we have used $\hbar\gamma=0.001 t, \mu/t=0.05$,  $\alpha = 0.5$, and $t/(k_BT) = 1000.$}
\label{sigma_lc}
\end{figure}
\\
\noindent $\dfrac{{\rm Re}[\sigma^{\eta\to 0}_{xx}]}{\sigma_0} =$
\begin{align}
\begin{cases}
    \dfrac{4 \pi\sqrt{\hbar\omega_{\rm pr} +2 \Delta }}{15\alpha\sqrt{t}\hbar^2\omega^2_{\rm pr}} \left[\left(4 \Delta ^2+3 \hbar^2\omega^2_{\rm pr}\right) {\rm Im}(E(\chi_p|\chi )-E(\chi ))\right.\\+\left.\hbar\omega_{\rm pr}  (2 \Delta +3 \hbar\omega_{\rm pr} ) {\rm Im}(K(\chi )-F(\chi_p|\chi ))\right], \hspace{0.15 cm}\hbar\omega_{\rm pr} < 2\Delta\\
    \\
    \dfrac{64 \pi \Delta^{3/2}}{15\omega_{\rm pr} \alpha  \sqrt{t}}, \hspace{4.12cm}\hbar\omega_{\rm pr} = 2\Delta\\
    \\
    \dfrac{4 \pi}{15\alpha\hbar^2\omega^2_{\rm pr}} \sqrt{\dfrac{\hbar\omega_{\rm pr} -2 \Delta }{t}}\left[\left(4 \Delta ^2+3 \hbar^2\omega^2_{\rm pr}\right) E\left(\frac{1}{\chi }\right)\right.\\+\left.\hbar\omega_{\rm pr}  (2 \Delta -3 \hbar\omega_{\rm pr} ) K\left(\frac{1}{\chi }\right)\right], \hspace{1.25cm}\hbar\omega_{\rm pr} > 2\Delta
\end{cases}\label{xx_analytical}
\end{align}
and,\\
\noindent $\dfrac{{\rm Re}[\sigma^{\eta\to 0}_{yy}]}{\sigma_0} =$
\begin{align}
\begin{cases}
    \dfrac{2\alpha \pi}{3\hbar^2\omega^2_{\rm pr}} \sqrt{\dfrac{t}{2 \Delta +\hbar\omega_{\rm pr} }}\left[\hbar\omega_{\rm pr}  (4 \Delta -\hbar\omega_{\rm pr}) {\rm Im}(F(\chi_p|\chi ))\right.\\-\left.4 \Delta  (2 \Delta +\hbar\omega_{\rm pr} ) {\rm Im}(E(\chi_p|\chi ))\right], \hspace{0.6cm} \hbar\omega_{\rm pr} < 2\Delta\\
    \\
  \dfrac{2\alpha \pi}{3\hbar^2\omega^2_{\rm pr}} \sqrt{\dfrac{t}{\hbar\omega_{\rm pr} -2\Delta}}\left[4 \Delta  (2 \Delta -\hbar\omega_{\rm pr} ) E\left(\frac{1}{\chi }\right)\right.\\+\left. \hbar\omega_{\rm pr}  (4 \Delta +\hbar\omega_{\rm pr} ) K\left(\frac{1}{\chi }\right)\right],\hspace{1.4cm} \hbar\omega_{\rm pr} > 2\Delta~,
\end{cases}\label{yy_analytical}
\end{align}
respectively, where we have defined $\sigma_0 = e^2/(\pi h)$, $\chi = (2 \Delta -\hbar\omega_{\rm pr} )/(2 \Delta +\hbar\omega_{\rm pr})$, $\chi_p = \sin^{-1}(\chi^{-1/2})$, $F(\chi_p|\chi )$ and $E(\chi_p|\chi )$ are elliptic integrals of the first and the second kind and $K(1/\chi)$ and $E(1/\chi)$ are the complete elliptic integrals of the first and the second kind respectively. The analytical expressions for the real parts of the conductivity match well with the numerical results obtained using Eq.~\eqref{conductivity_Kubo} as can be seen from Fig. \ref{sigma_lc}. These results are calculated for positive $\Delta$ at zero temperature as well as zero doping level. For finite temperature and non-vanishing chemical potential, expressions in Eq.~\eqref{xx_analytical} and Eq.~\eqref{yy_analytical} needs to be multiplied with a factor,
\begin{align}
g(\omega_{\rm pr},\mu,T) = \frac{1}{2}\sum_{s=\pm}\tanh\left(\frac{\hbar\omega_{\rm pr}+2s\mu}{4k_BT}\right) ~.
\end{align}
%
As $\hbar\omega_{\rm pr}\to 2\Delta$, $\sigma_{yy}$ diverges whereas $\sigma_{xx}$ remains finite. Similarly, we can perform the calculation for the $-\Delta$ case. Finally, in the linear regime, the transverse optical conductivity vanishes.
\subsection{Nonlinear regime}
In the nonlinear regime, the population of the conduction and the valence band is modified from the equilibrium thermal distribution. It is clear from Eq. \eqref{n_k}, that the difference between the non-equilibrium population distribution between the two bands, $n_{\bf k}$ decreases as $\eta$ increases.  Therefore, the optical conductivity is expected to decrease (as compared to the linear response case) w.r.t. $\eta$. Before we discuss the more general scenario, we illustrate this analytically in the small $\eta$ limit.
\begin{figure}[ht!]
\includegraphics[width =0.85\linewidth]{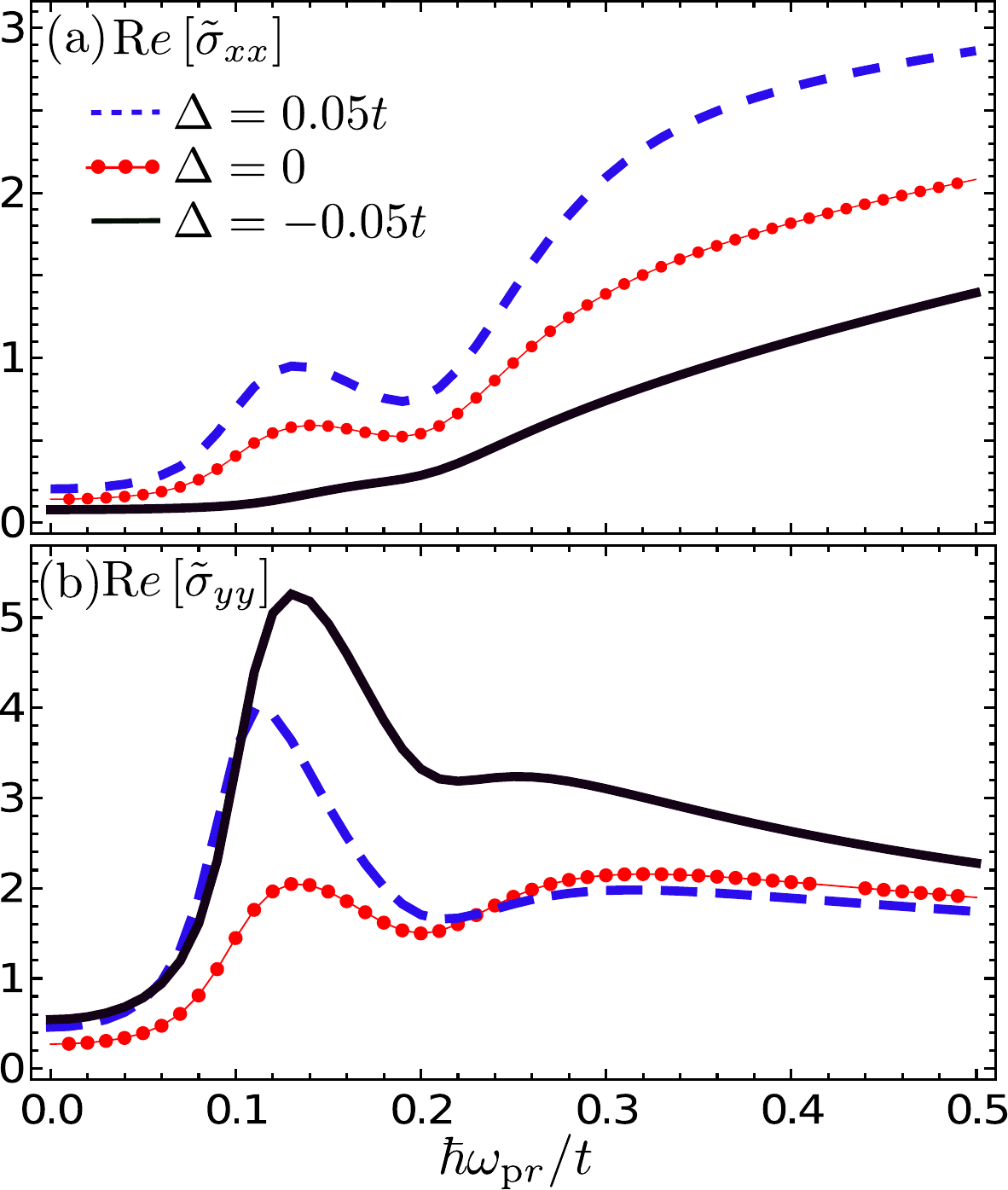}
\caption{Real parts of normalized optical conductivity ($\tilde\sigma_{ii}\equiv  \sigma_{ii}/\sigma_0$) as a function of the normalized probe photon energy for positive, negative, and zero $\Delta$. Here $\sigma_0 = e^2/(\pi h), \omega = 0.2t/\hbar, \theta = 0, \hbar\gamma_2 = \hbar\gamma=0.02 t, \mu/t=0.05$,  $\alpha = 0.5$, $\eta = 10$, and $t/(k_BT) = 300$. The energy cut-off for the momentum integration is set to be $\hbar\omega_c = 5 t$.}
\label{sigma}
\end{figure}
For small values of $\eta$, keeping terms upto second order in $\eta$ yields,
 \begin{equation}
n^{\eta^2}_{\bf k} =n_{\bf k}\big{|}_{\eta< 1}\approx  n^{\rm eq}_{\bf k}\left[1-\eta^2\frac{\omega^2}{\omega^2_{\bf k}}\frac{|{\mathcal M}^{vc}_{\bf k}\cdot \hat {\bf e}_0|^2 }{(eat/\hbar)^2} \sum_{s=\pm} {\mathcal W}_s(\omega_{\bf k},\omega)\right]~.
 \end{equation}
In order to compute the conductivity analytically, we take $\omega_{\rm pr}\gg\gamma$ limit, such that in the small $\eta$ limit we have,
\begin{align}
{\rm Re}[\sigma^{\eta^2}_{ij}(\omega_{\rm pr};\omega)] = -g_s\int \frac{d{\bf k}}{(2\pi)^2}\frac{ n^{\eta^2}_{\bf k}}{\hbar\omega_{\bf k}}{\mathcal M}^{vc}_{{\bf k}i}{\mathcal M}^{cv}_{{\bf k}j}\pi\delta(\omega_{\bf k} -\omega_{\rm pr})~.
\end{align}
Since the analytical expressions in the linear regime for finite $\Delta$ cases are lengthy and complicated, in the present case we only focus on the $\Delta = 0$ case. The expression for the real parts of the $xx$-, $yy$-  and $xy$- components of nonlinear conductivities in the leading order in $\eta$ are given as,
\begin{widetext}
\begin{eqnarray}
    {\rm Re}[\sigma^{\eta^2}_{xx}(\omega_{\rm pr};\omega)] &=& \frac{2e^2}{h}\Bigg[\dfrac{5 \sqrt{\tilde{\omega}_{\rm pr}} \Gamma[-\frac{5}{4}]^2}{128 \sqrt{2 \pi}\alpha}-\eta^2\frac{\omega^2}{\omega^2_{\rm pr}}\mathcal{P}(\tilde{\omega}_{\rm pr},\theta)\sum_{s=\pm} {\mathcal W}_{s}(\tilde\omega_{\rm pr},\tilde\omega)
    \Bigg] g(\tilde{\omega}_{\rm pr},\tilde\mu,T)\label{xx_eta_sq}~,\\
     {\rm Re}[\sigma^{\eta^2}_{yy}(\omega_{\rm pr};\omega)]  &=&  \frac{2e^2}{h}\Bigg[\dfrac{ \alpha\Gamma[\frac{1}{4}]^2}{12 \sqrt{\tilde{\omega}_{\rm pr}}\sqrt{2 \pi}}-\eta^2\frac{\omega^2}{\omega^2_{\rm pr}}\mathcal{Q}(\tilde{\omega}_{\rm pr},\theta)\sum_{s=\pm} {\mathcal W}_{s}(\tilde\omega_{\rm pr},\tilde\omega)
    \Bigg]g(\tilde{\omega}_{\rm pr},\tilde\mu,T)\label{yy_eta_sq}~,\\
     {\rm Re}[\sigma^{\eta^2}_{xy}(\omega_{\rm pr};\omega)] &=& - \frac{2e^2}{h}\alpha\eta^2\frac{\omega^2}{\omega^2_{\rm pr}} \dfrac{\sqrt{2} \Gamma[\frac{3}{4}]^2 \tilde{\omega}_{\rm pr}^{1/2} \sin{2 \theta}}{15 \sqrt{ \pi}}
     \sum_{s=\pm} {\mathcal W}_{s}(\tilde\omega_{\rm pr},\tilde\omega)g(\tilde{\omega}_{\rm pr},\tilde\mu,T)\label{xy_eta_sq}~,
     \end{eqnarray}
\end{widetext}
where we have defined,
\begin{align}
    \mathcal{P}(\tilde{\omega}_{\rm pr},\theta)=\dfrac{\alpha\sqrt{2\tilde{\omega}_{\rm pr}}\Gamma[\frac{3}{4}]^2 \sin^2{\theta}}{15\sqrt{\pi}}
    +\dfrac{63 \Gamma[-\frac{7}{4}]^2 \tilde{\omega}_{\rm pr}^{3/2} \cos^2{\theta}}{2816 \sqrt{2 \pi}\alpha^2}\\
    \mathcal{Q}(\tilde{\omega}_{\rm pr},\theta)=\dfrac{15 \alpha^{3} \Gamma[-\frac{3}{4}]^2 \sin^2{\theta}}{448\sqrt{2 \tilde{\omega}_{\rm pr} \pi}}
    +\dfrac{2 \alpha \Gamma[\frac{3}{4}]^2 \tilde{\omega}_{\rm pr}^{1/2} \cos^2{\theta}}{15 \sqrt{2 \pi}}
\end{align}
Here, $\eta$ independent terms exist only in the longitudinal components such that the overall magnitude decreases with the increase in $\eta$. This is in contrast to the case of the transverse conductivity, which is only supported by the finite $\eta$.
\begin{figure}[t]
\includegraphics[width =0.9\linewidth]{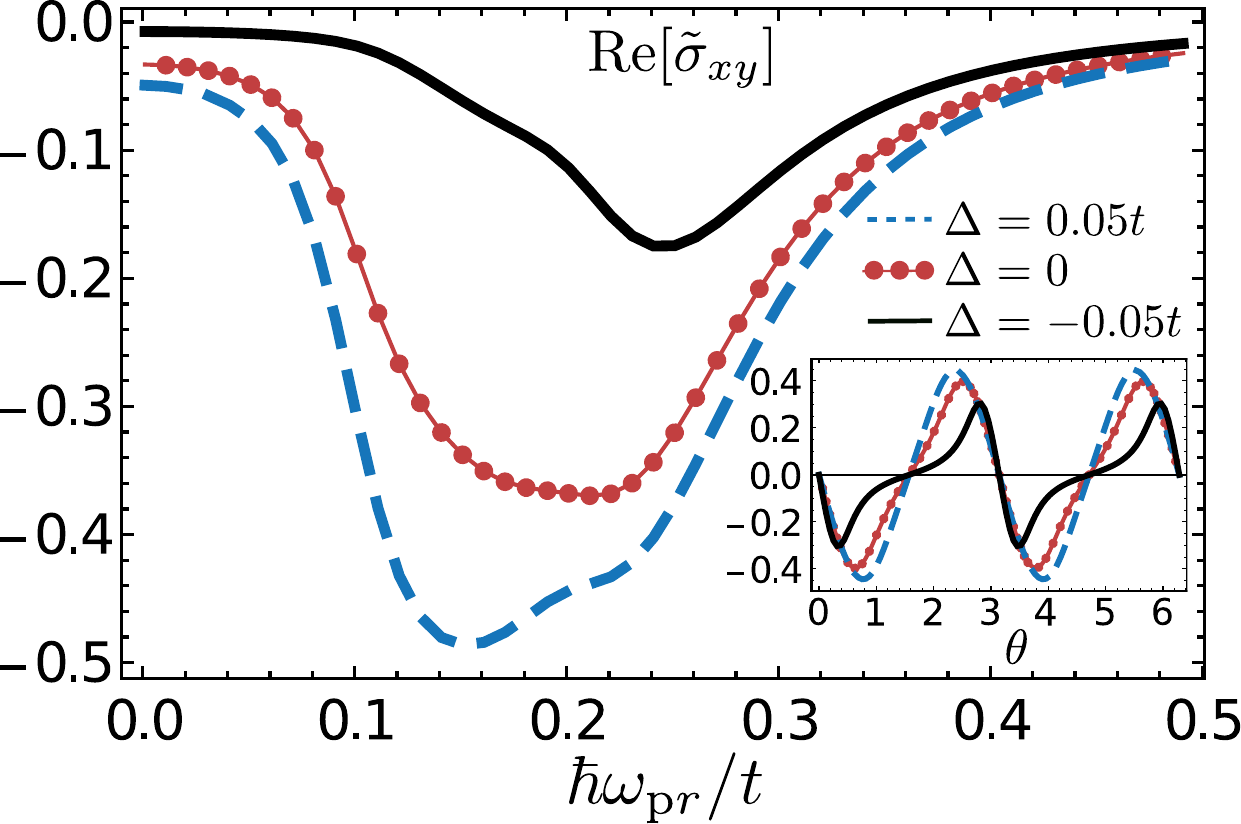}
\caption{Real part of normalized optical conductivity ($\tilde\sigma_{xy}\equiv  \sigma_{xy}/\sigma_0$) as a function of the normalized probe photon energy with a fixed value of polarization angle $( \theta = \pi/4)$ for positive, negative, and zero $\Delta$. (Inset) Variation of $\tilde\sigma_{xy}$ with $\theta$, keeping a fixed value of probe photon energy $(\hbar \omega_{\rm pr} =0.2t)$. Other parameters are same as in Fig.\ref{sigma}.}
\label{sigma_xy}
\end{figure}
Full numerical results based on the expression given in Eq.~\eqref{conductivity} are shown in Fig.~\ref{sigma} and Fig.~\ref{sigma_xy}. Due to the finite $\gamma_2$ and temperature-led broadening, the conductivities remain small, but non-zero for $\omega\to 0$. In principle, there exists a large Drude contribution in low-frequency regions, which we have not included in our discussion due to its contribution being insignificant in high-frequency regions. The peak in the $yy$- component attributes its origin to the diverging DOS (Fig.~\ref{DOS}), which is softened due to the presence of decay term and temperature effects. No such peak exists in the $xx$- component due to the different structure of the transition matrix element. At high frequencies, the conductivities tend to follow the frequency dependence derived analytically. The effect of the pump field is obvious in the region where $\omega_{\rm pr}\approx \omega$. The transverse part of the conductivity vanishes both in the low as well as high frequency regions, and remains finite in the vicinity of pump frequency. In contrast to the longitudinal parts, the transverse conductivity changes sign w.r.t. the change in polarization angle $\theta$, as shown in the inset (Fig.~\ref{sigma_xy}).
\subsection{Transmission coefficient}\label{sec:Transmission}
Experimental data for optical conductivity is usually fetched from the transmission spectra of the sample\cite{Tanner_2019}, which in reality can become a very complicated task, especially when the nonlinear response of the system under investigation is concerned. Further complications arise in two-dimensional systems which illustrate small absorbance, e.g. in graphene, it is approximately $\pi/137$ (Ref. \cite{PhysRevLett.101.196405}).
However, unlike graphene, the low-energy dispersion in a semi-Dirac system is highly anisotropic.
Since the longitudinal optical conductivities are not equal in this case, it is expected that transmission will display strong polarization dependence. In this section, we compute the transmission coefficients, which are related to the $s$- and $p$- components of the incident and the transmitted waves\cite{PhysRevB.97.205420},
 %
\begin{align}\label{eq:tss}
t_{ss} = \left. \frac{E_t^s}{E_i^s}\right|_{E_i^p = 0} = \frac{2n_i}{c\mu_0}\frac{\sigma_2}{\sigma_T}\cos\theta_i,
\end{align}
and,
\begin{align}\label{eq:tpp}
t_{pp} = \left. \frac{E_t^p}{E_i^p}\right|_{E_i^s = 0} = \frac{2n_i}{c\mu_0}\frac{\sigma_1}{\sigma_T}\cos\theta_i,
\end{align}
where $E_i^{\rm pol}, E_t^{\rm pol}$ (${\rm pol}\in [s,p]$) are the incident and the transmitted components of the electric field vector, $n_i(n_t)$ is the refractive index of the medium of incidence(transmission), $\mu_0$ is free permeability, $\theta_i$ and $\theta_t$ are the angle of incidence and transmission, and
\begin{eqnarray}
\sigma_T &=& \left(\sigma_1\sigma_2 - \sigma_{xy}\sigma_{yx}\cos\theta_i\cos\theta_t \right), \\
\sigma_1 &=& n_i\cos\theta_i/(c\mu_0) + n_t\cos\theta_t/(c\mu_0) + \sigma_{xx}, \\\nonumber
\sigma_2 &=& n_i\cos\theta_t/(c\mu_0) + n_t\cos\theta_i/(c\mu_0) + \sigma_{yy}\cos\theta_i\cos\theta_t~.\\
\end{eqnarray}
Furthermore, we can have finite transverse conductivity due to the non-equilibrium population distribution. The off diagonal part of the transmission matrix that is related to $\sigma_{xy}$, is given by,
\begin{align}\label{eq:tsp}
t_{sp} = \left. \frac{E_t^s}{E_i^p}\right|_{E_i^s = 0} = -\frac{2n_i}{c\mu_0}\frac{\sigma_{xy}}{\sigma_T}\cos\theta_i\cos\theta_t,
\end{align}
%
\begin{figure}[t!]
\includegraphics[width =\linewidth]{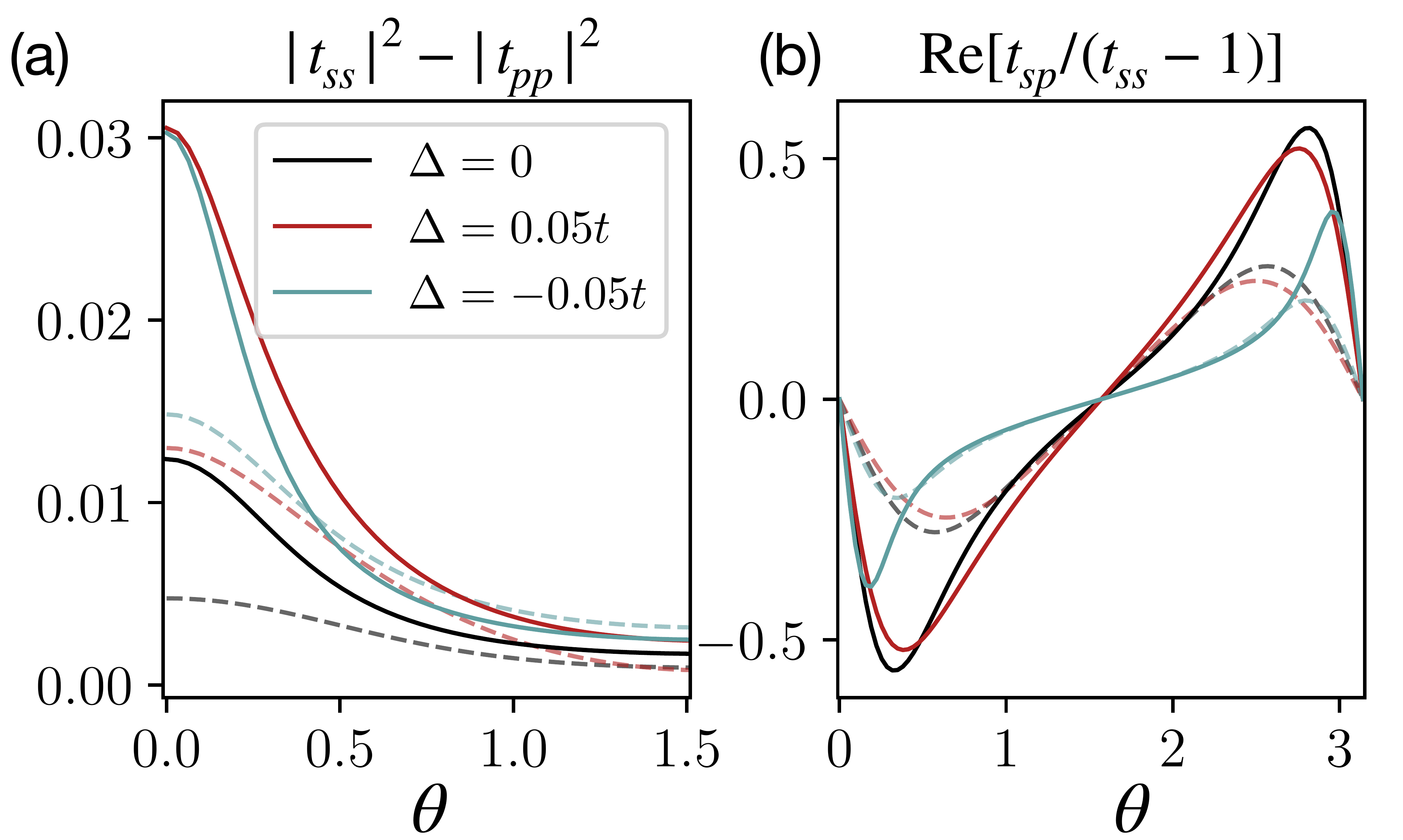}
\caption{(a). Difference between $s$- and $p$- transmittance w.r.t. polarization angle ($\theta$) and (b) Kerr angle variation w.r.t. $\theta$ for three values of $\Delta$. The parameters used here are $\hbar\omega_{\rm pr} = 0.1 t$, $\alpha = 1$ ($1/2$ for dashed lines), $\theta_i = 0$ and $\theta_t = 0$. Other parameters are same as in Fig.~\ref{sigma}.}
\label{Transmission}
\end{figure}
In Fig. \ref{Transmission}, we plot the variation of the difference between the transmittance corresponding to $s$- and $p$- components and Kerr angle w.r.t. $\theta$ for the normal angle of incidence. For comparison, we have considered three different values of gap parameter $\Delta$.
Since transverse conductivity components are much smaller in magnitude as compared to the longitudinal components (Eq.\eqref{xx_eta_sq}-Eq.\eqref{xy_eta_sq}), we can approximate $\sigma_T \approx \sigma_1\sigma_2$. This immediately implies that $t_{ss}$ and $t_{pp}$ approximately follows
$(1 + c\mu_0\sigma_{xx}/2)^{-1}$ and $(1 + c\mu_0\sigma_{yy}/2)^{-1}$ respectively. From Fig. \ref{sigma} we know that $\sigma_{xx}$ and $\sigma_{yy}$ are within a few order of $\sigma_0$, suggesting that $\sigma_{xx/yy} \ll 2/(c\mu_0)$. Therefore, we have $t_{ss/pp}\approx 1 - c\mu_0\sigma_{xx/yy}/2$.
For pump polarization angle $\theta = 0$ and probe photon energy $\hbar\omega_{\rm pr} = 0.1 t$, $\sigma_{yy}$ values are roughly equal for positive and negative $\Delta$ whereas $\sigma_{xx}$ values remain smaller than 1. This is reflected in the plots for $|t_{ss}|^2-|t_{pp}|^2$ (Fig. \ref{Transmission}(a)). For polarization angles $\theta\neq n\pi, n = 0,1...,$ anisotropic effects are reduced. Furthermore, the real part of $t_{sp}/(t_{ss}-1)$, which refers to the Kerr angle, is mainly dictated by the real part of $\sigma_{xy}$. This can be confirmed from Eq.~\eqref{xy_eta_sq} and the polarization variation of the Kerr angle as shown in Fig. \ref{Transmission}(b). More dramatic results are expected for finite angle of incidence, $\theta_i$.
\section{Conclusions} 
In this paper, we have studied the nonlinear response for the Hamiltonian displaying semi-Dirac electron dispersion, which demonstrates parabolic as well as linear dependence on momentum in the $x$- and the $y$- direction respectively. To describe the non-equilibrium carrier dynamics, we use the equation of motion approach of the density matrix. Under CW illumination of the pump field, the competition between the carrier excitation from the valance to the conduction band and the subsequent relaxation results in a quasi-steady state of the carrier population in the conduction band. The non-equilibrium carrier distribution thus obtained, depends on the parameters of the pump field, in contrast to the Fermi-Dirac distribution of the carriers. Accordingly, optical conductivity corresponding to a probe field strongly depends on the electric field strength, frequency, and the polarization direction of the pump field. Moreover, optical conductivity for the probe field shows a minimum whenever the probe frequency coincides with that of the pump. The dip in the inter-band optical conductivity spectra is due to the spectral weight transfer from the inter-band to the intra-band regime since the total spectral weight is conserved. These nonlinear features can be controlled using a dimensionless parameter $\eta$, such that $\eta\sim\mathcal{O}(1)$ or larger leads to a significant departure from the equilibrium response. 

Furthermore, the non-equilibrium carrier distribution also gives rise to transverse optical conductivity, which is in contrast to the equilibrium situation. Indeed, the analytical results obtained in the small $\eta$ limit confirm that $\sigma_{xy}$ vanishes as $\eta^2$. We find that the polarization rotation as seen in the transmission profile reaches up to $\sim 0.5$ for $\eta = 10$ (Fig.~\ref{Transmission}(b)), which corresponds to electric field $E_0 = \hbar^2\omega\sqrt{\gamma_1\gamma_2}/(eat)\approx 10^8$Vm$^{-1}$ for parameters in Fig.~\ref{sigma}, and assuming $\gamma_1 = \gamma_2/10$, $a\sim 1 \mathring{A}$ and $t\sim1$ eV. Note that $\gamma_1$ and $\gamma_2$ represent the scattering effects such as carrier-carrier and carrier-phonon interactions and hence depend on the momentum, and temperature as well as the pump frequency at which the carriers are excited from the valance band to the conduction band. 

Another important feature of using optical pumping in this system is that the default anisotropic response ($\sigma^{\eta\to 0}_{yy}\neq\sigma^{\eta\to 0}_{xx}$) can be manipulated and controlled, since $\sigma_{yy}/\sigma_{xx}$ can be larger or smaller than $\sigma^{\eta\to 0}_{yy}/\sigma^{\eta\to 0}_{xx}$, depending upon the pump field parameters. This can lead to several practical consequences such as tunable reflection and transmission profiles, as well as giving rise to a controllable frequency window over which the system displays hyperbolicity\cite{PhysRevApplied.12.014011, PhysRevLett.131.096902}.
Exploring these features in realistic systems, e.g. in nodal-line metal ZrSiS\cite{shao2023semidiracfermionstopologicalmetal} can be interesting since the low-energy quasiparticles in such nodal line systems can be described using a semi-Dirac dispersion.

\section{ACKNOWLEDGEMENTS}
We would like to thank Alexey Belyanin for illuminating discussions. BG is supported by INSPIRE, DST, Government of India. MB is supported by the Department of Science and Technology (DST), Government of India under the Core grant (Project No. CRG/2020//001768) and MATRICS grant (Project no. MTR/2021/000566). AS is supported in part by the Air Force Office for Scientific Research Grant No. FA9550- 21-1-0272 and National Science Foundation Award No. 1936276.
\appendix\label{Sec:appendix}
\section{Inter-band optical conductivity: Real part}
In this section we provide analytical calculation for the real part of the optical conductivity upto $\eta^2$ for $\Delta = 0$ case. 
\subsection{Linear regime}
The real part of the interband conductivity can be calculated using Eq.(\ref{conductivity}),
\begin{align}\label{A1}
\frac{{\rm Re}[\sigma_{ij}(\omega_{\rm pr};\omega)]}{
    {g_s}/{(4\pi^{2}\hbar)}}=&- \int \frac{d{\bf k}}{\omega_{\bf k}} {\mathcal M}^{vc}_{{\bf k}i}{\mathcal M}^{cv}_{{\bf k}j} n_{\bf k} \sum_{s=\pm} \frac{{\mathcal W}_s(\omega_{\bf k},\omega_{\rm pr})}{\gamma_2}\bigg{|}_{\gamma_2\to \gamma}~.
\end{align}
For $\eta\to 0$ (linear regime), $n_{\bf k} \to n^{\rm eq}_{\bf k}$. Assuming $\omega_{\rm pr}\gg\gamma$, the Lorentzian becomes delta function such that,
\begin{align}\label{A2}
    {\rm Re}&[\sigma^{\eta\to 0}_{xx}(\omega_{\rm pr})]\nonumber\\
    &=-\frac
    {g_s}{4\pi^{2}\hbar} \int \frac{d{\bf k}}{\omega_{\bf k}}|{\mathcal M}^{vc}_{{\bf k}x}|^2 n^{\rm eq}_{\bf k} \frac{\gamma}{\gamma^2 + (\omega_{\bf k}-\omega_{\rm pr})^2}~,\nonumber\\
    &=-\dfrac{ e^2}{h}\int \frac{4 \tilde k^2_x \tilde k^2_y \alpha^2}{\tilde k^4_x + 4 \alpha^2 \tilde k^2_y} n^{\rm eq}_{\bf k} \delta(\tilde \omega_{\bf k}-\tilde {\omega}_{\rm pr}) \frac{\tilde k ~d \tilde k }{\tilde \omega_{\bf k}}d\phi_{\bf k}~,
\end{align}
where $\tilde k_x = k_x a, \tilde k_y = k_y a, \tilde\Delta=\Delta/t, \tilde\omega_{\bf k} = \hbar\omega_{\bf k}/t$ and  $\tilde\omega_{\rm pr} = \hbar\omega_{\rm pr}/t.$
By substuting $\tilde k_x= \tilde k \cos{\phi_{\bf k}}$ and $\tilde k_y=\tilde k \sin{\phi_{\bf k}}$ in Eq.(\ref{A2}), we obtain
%
\begin{align}
       &{\rm Re}[\sigma^{\eta\to 0}_{xx}( \omega_{\rm pr})]\nonumber\\
       &=-\dfrac{ e^2 \alpha^2}{h }\int \frac{n^{\rm eq}_{\bf k}4 \tilde k^2 \cos^2{\phi_{\bf k}} \sin^2{\phi_{\bf k}}\delta(\tilde \omega_{\bf k}-\tilde {\omega}_{\rm pr})}{ \tilde k^2 \cos^4{\phi_{\bf k}} + 4 \alpha^2 \sin^2{\phi_{\bf k}}}  \frac{\tilde k ~d \tilde k }{\tilde \omega_{\bf k}}d\phi_{\bf k}
\end{align}
Here,
$\hbar \omega_{\bf k}= t (\tilde k^4_x + 4 \alpha^2 \tilde k^2_y)^{1/2}$ and $\tilde \omega_{\bf k}=(\tilde k^4_x + 4 \alpha^2 \tilde k^2_y)^{1/2}$.
From this we can get $\tilde k^2$ and $\tilde k~ \tilde dk$
in terms of $\tilde \omega_{\bf k}$. We have, 
\begin{equation}
    \begin{aligned}
        \tilde k^{2}&=\frac{-2 \alpha^2 \sin^2{\phi_{\bf k}}+S}{\cos^4{\phi_{\bf k}}},\\
    \end{aligned}
\end{equation}
and
\begin{equation}
    \begin{aligned}
        \tilde k~\tilde dk&=\frac{\tilde \omega_{\bf k} ~d \tilde \omega_{\bf k} }{2S},\\
    \end{aligned}
\end{equation}
where $S= (X^2 + \tilde \omega^2_{\bf k} \cos^4{\phi_{\bf k}} )^{1/2}$ and $X=2 \alpha^2 \sin^2{\phi_{\bf k}}$. 
So,
\begin{align}
    &{\rm Re}[\sigma^{\eta\to 0}_{xx}(\omega_{\rm pr})]\nonumber\\
    &=-\dfrac{ 2 e^2  \alpha^2}{h}\int \frac{ \tan^2{\phi_{\bf k}} (S-X)}{ S (S+X)} n^{\rm eq}_{\bf k}\delta(\tilde \omega_{\bf k}-\tilde {\omega}_{\rm pr})d\tilde{\omega}_{\bf k}~ d\phi_{\bf k}~, \nonumber\\
    &=\dfrac{ e^2}{h} \dfrac{5 \sqrt{\tilde{\omega}_{\rm pr}} \Gamma[-\frac{5}{4}]^2}{64 \sqrt{2 \pi}\alpha}  g(\tilde{\omega}_{\rm pr},\tilde\mu,\tilde T)~.
\end{align}
Similarly, the real part of $\sigma_{yy}$ can be calculated as,    
\begin{align}
    &{\rm Re}[\sigma^{\eta\to 0}_{yy}( \omega_{\rm pr})]\nonumber\\
    & =-\frac
    {g_s}{(2\pi)^{2}\hbar} \int |{\mathcal M}^{vc}_{{\bf k}y}|^2 n^{\rm eq}_{\bf k} \frac{\gamma}{\gamma^2 + (\omega_{\bf k}-\omega_{\rm pr})^2} \frac{d{\bf k}}{\omega_{\bf k}}~,\nonumber\\
    &=-\dfrac{ e^2}{h}\int \frac{k^2 \alpha^2 \cos^4{\phi_{\bf k}}}{ \tilde k^2 \cos^4{\phi_{\bf k}} + 4 \alpha^2 \sin^2{\phi_{\bf k}}} n^{\rm eq}_{\bf k} \delta(\tilde \omega_{\bf k}-\tilde {\omega}_{\rm pr}) \frac{\tilde k ~d \tilde k }{\tilde \omega_{\bf k}}d\phi_{\bf k}~,\nonumber\\
        &=-\dfrac{ e^2  \alpha^2}{2 h}\int \frac{  (S-X)}{ S (S+X)} n^{\rm eq}_{\bf k} \delta(\tilde \omega_{\bf k}-\tilde {\omega}_{\rm pr}) d\tilde{\omega}_{\bf k} ~d\phi_{\bf k}~,\nonumber\\
        &=\dfrac{ e^2}{ h}\dfrac{ \Gamma[\frac{1}{4}]^2 \alpha}{6\sqrt{\tilde{\omega}_{\rm pr}}\sqrt{2 \pi}} g(\tilde{\omega}_{\rm pr},\tilde\mu,\tilde T)~.
\end{align}
Finally, the transverse component of the inter-band optical conductivity can be shown to vanish in the linear regime,
\begin{align}    
    &{\rm Re}[\sigma^{\eta\to 0}_{xy}( \omega_{\rm pr})]\nonumber\\
    &=- \frac
    {g_s}{(2\pi)^{2}\hbar} \int {\mathcal M}^{vc}_{{\bf k}x} {\mathcal M}^{cv}_{{\bf k}y} n^{\rm eq}_{\bf k} \frac{\gamma}{\gamma^2 + (\omega_{\bf k}-\omega_{\rm pr})^2}\frac{d{\bf k}}{\omega_{\bf k}}=0~.
\end{align}
\subsection{Nonlinear regime}
In this case, we analytically calculate the contribution to the optical conductivity upto $\eta^2$. In the $\eta^2\ll1$ limit, $n_{\bf k}$ becomes,
 \begin{equation}\label{A10}
 n^{\eta^2}_{\bf{k}}= n^{\rm eq}_{\bf k}  \left(1-\eta^2\frac{\omega^2}{\omega^2_{\bf k}}\frac{|{\mathcal M}^{vc}_{\bf k}\cdot \hat {\bf e}_0|^2 }{\mathcal{V}^2} \sum_{s=\pm} {\mathcal W}_{s}(\omega_{\bf{k}},\omega)\right)~,
 \end{equation}
where, 
\begin{equation}
    \begin{aligned}
        |{\mathcal M}^{vc}_{\bf k}\cdot{\bf e}_0|^2&=|{\mathcal M}^{vc}_{{\bf k}x}\cos{\theta}+ {\mathcal M}^{vc}_{{\bf k}y}\sin{\theta}|^2,\\
        &=\frac{ \mathcal{V}^2 \alpha^2 \tilde k^{2}_x(\tilde k_x\sin{\theta}-2 \tilde k_y \cos{\theta})^2}{(\tilde k^4_x + 4 \alpha^2 \tilde k^2_y)},\\
        &=\frac{\mathcal{V}^2 \alpha^2 \tilde k^2 \cos^2{\phi_{\bf k}} (\cos{\phi_{\bf k}} \sin{\theta}-2 \cos{\theta} \sin{\phi_{\bf k}})^2}{ (\tilde k^2 \cos^4{\phi_{\bf k}} + 4 \alpha^2 \sin^2{\phi_{\bf k}})},\\
        &=\frac{\alpha^2 \mathcal{V}^2 (S-X) \mathcal{F}(\theta,\phi_{\bf k})}{(S+X)}~,
    \end{aligned}
\end{equation}
$\mathcal{F}(\theta,\phi_{\bf k})=( \sin{\theta}-2 \cos{\theta} \tan{\phi_{\bf k}})^2$ and $\mathcal{V}=eat/\hbar$.
To compute the real part of interband conductivity up to the second order of $\eta$, we utilize Eq.(\ref{A1}) and substitute $n_{\mathbf{k}}$ by $n^{\eta^2}_{\bf{k}}$, given in Eq.(\ref{A10}). The resulting $xx$ component of the conductivity is given by,

\begin{align}
    &{\rm Re}[\sigma^{\eta^2}_{xx}( \omega_{\rm pr};\omega)]\nonumber\\
    &= -\frac
    {g_s}{(2\pi)^{2}\hbar} \int |{\mathcal M}^{vc}_{{\bf k}x}|^2 n^{\eta^2}_{\bf{k}} \frac{\gamma}{\gamma^2 + (\omega_{\bf k}-\omega_{\rm pr})^2}\frac{d{\bf k}}{\omega_{\bf k}}~,\nonumber\\
    &={\rm Re}[\sigma^{\eta\to 0}_{xx}(\tilde \omega_{\rm pr})]+ \dfrac{2 e^2}{h} \eta^2 \alpha^4 \frac{\tilde \omega^2}{\tilde \omega^2_{\rm pr}}\sum_{s=\pm} {\mathcal W}_{s}(\tilde \omega_{\rm pr},\tilde \omega)\times\nonumber\\
    &\qquad\int \frac{  \tan^2{\phi_{\bf k}} (S-X)^2n^{\rm eq}_{\bf k} \mathcal{F}(\theta,\phi_{\bf k})}{ S (S+X)^2} d\phi_{\bf k}~,\nonumber\\
    &=\frac{2e^2}{h}\Bigg[\dfrac{5 \sqrt{\tilde{\omega}_{\rm pr}} \Gamma[-\frac{5}{4}]^2}{128 \sqrt{2 \pi}\alpha} - \eta^2\frac{\tilde \omega^2}{\tilde \omega^2_{\rm pr}}\mathcal{P}(\tilde{\omega}_{\rm pr},\theta)\times \nonumber\\
    &\hspace{2.5cm}\sum_{s=\pm} {\mathcal W}_{s}(\tilde \omega_{\rm pr},\tilde \omega)
    \Bigg] g(\tilde{\omega}_{\rm pr},\tilde\mu,T)~,
\end{align}
where,
\begin{equation}
    \mathcal{P}(\tilde{\omega}_{\rm pr},\theta)=\dfrac{\alpha\sqrt{2\tilde{\omega}_{\rm pr}}\Gamma[\frac{3}{4}]^2 \sin^2{\theta}}{15\sqrt{\pi}}
    +\dfrac{63 \Gamma[-\frac{7}{4}]^2 \tilde{\omega}_{\rm pr}^{3/2} \cos^2{\theta}}{2816 \sqrt{2 \pi}\alpha^2}~.
\end{equation}
Similarly, the real part of the $yy$ component of the  interband nonlinear optical conductivity is given by,
\begin{align}
    &{\rm Re}[\sigma^{\eta^2}_{yy}( \omega_{\rm pr}; \omega)]\nonumber\\
    &= -\frac
    {g_s}{(2\pi)^{2}\hbar} \int |{\mathcal M}^{vc}_{{\bf k}y}|^2 n^{\eta^2}_{\bf{k}} \frac{\gamma}{\gamma^2 + (\omega_{\bf k}-\omega_{\rm pr})^2}\frac{d{\bf k}}{\omega_{\bf k}}~,\nonumber\\
    &={\rm Re}[\sigma^{\eta\to 0}_{yy}(\tilde \omega_{\rm pr})]+\dfrac{ e^2}{2 h} \eta^2 \alpha^4 \frac{\tilde \omega^2}{\tilde \omega^2_{\rm pr}}\sum_{s=\pm} {\mathcal W}_{s}(\tilde \omega_{\rm pr},\tilde \omega)\times\nonumber\\
    &\qquad\qquad \int \frac{   (S-X)^2 n^{\rm{e q}}_{\bf k}\mathcal{F}(\theta,\phi_{\bf k})}{ S (S+X)^2} d\phi_{\bf k}~,\nonumber\\
    &=\frac{2e^2}{h}\Bigg[\dfrac{ \alpha\Gamma[\frac{1}{4}]^2}{12 \sqrt{\tilde{\omega}_{\rm pr}}\sqrt{2 \pi}}-\eta^2\frac{\tilde \omega^2}{\tilde \omega^2_{\rm pr}}\mathcal{Q}(\tilde{\omega}_{\rm pr},\theta)\times\nonumber\\
    &\hspace{2.5cm} \sum_{s=\pm} {\mathcal W}_{s}(\tilde \omega_{\rm pr},\tilde \omega)
    \Bigg]g(\tilde{\omega}_{\rm pr},\tilde\mu,\tilde T)~,
\end{align}
where,
\begin{equation}
    \mathcal{Q}(\tilde{\omega}_{\rm pr},\theta)=\dfrac{15 \alpha^{3} \Gamma[-\frac{3}{4}]^2 \sin^2{\theta}}{448\sqrt{2 \tilde{\omega}_{\rm pr} \pi}}
    +\dfrac{2 \alpha \Gamma[\frac{3}{4}]^2 \tilde{\omega}_{\rm pr}^{1/2} \cos^2{\theta}}{15 \sqrt{2 \pi}}~.
\end{equation}
In contrast to the linear regime, the transverse conductivity does not vanish in the nonlinear regime. We have,
%
\begin{align}   
    &{\rm Re}[\sigma^{\eta^2}_{xy}( \omega_{\rm pr}; \omega)]\nonumber\\
    &= -\frac
    {g_s}{(2\pi)^{2}\hbar} \int {\mathcal M}^{vc}_{{\bf k}x} {\mathcal M}^{cv}_{{\bf k}y} n^{\eta^2}_{\bf{k}}  \frac{\gamma}{\gamma^2 + (\omega_{\bf k}-\omega_{\rm pr})^2}\frac{d{\bf k}}{\omega_{\bf k}}~,\nonumber \\  
    &=-\dfrac{ e^2}{ h} \eta^2 \alpha^4 \frac{\tilde \omega^2}{\tilde \omega^2_{\rm pr}}\sum_{s=\pm} {\mathcal W}_{s}(\tilde \omega_{\rm pr},\tilde \omega)\times\nonumber\\
    &\qquad\qquad\int \frac{ \tan{\phi_{\bf k}} (S-X)^2 n^{\rm{e q}}_{\bf k}\mathcal{F}(\theta,\phi_{\bf k})}{ S (S+X)^2} d\phi_{\bf k}~,\nonumber\\
    &=- \frac{2e^2}{h}\eta^2\frac{\tilde \omega^2}{\tilde \omega^2_{\rm pr}} \dfrac{\sqrt{2} \alpha \Gamma[\frac{3}{4}]^2 \tilde{\omega}_{\rm pr}^{1/2} \sin{2 \theta}}{15 \sqrt{ \pi}}\times\nonumber\\
    &\hspace{2.5cm}\sum_{s=\pm} {\mathcal W}_{s}(\tilde \omega_{\rm pr},\tilde \omega)g(\tilde{\omega}_{\rm pr},\tilde\mu,T)~.
\end{align}
\section{Inter-band optical conductivity: Imaginary part}
Now we can calculate imaginary part of the conductivity from real part of the conductivity by using the Kramers-Kronig relation which is defined by,
\begin{equation}
    {\rm Im}[\sigma_{ij}( \omega_{\rm pr}; \omega)]=\dfrac{2 \omega_{\rm pr}}{\pi} \mathcal{P} \int^{\infty}_{0} \dfrac{{\rm Re}[\sigma_{ij}( \omega^{\prime}_{\rm pr}; \omega)]}{\omega_{\rm pr}^2 -\omega^{\prime ~2}_{\rm pr}} d\omega^{\prime}_{\rm pr}~.
\end{equation}
\subsection{Linear regime}
For $T = 0$, $g(\omega_{\rm pr},\mu,T)=\Theta\left(\dfrac{\hbar\omega_{\rm pr}}{2}-|\mu|\right)$.
The imaginary parts of the $xx$ and $yy$ components of the optical conductivity in linear regime are given by, 
\begin{align}
     &{\rm Im}[\sigma^{\eta\to 0}_{xx}( \omega_{\rm pr})]\nonumber\\
     &=\dfrac{2 \omega_{\rm pr}}{\pi} \mathcal{P} \int^{\infty}_{0} \dfrac{{\rm Re}[\sigma^{\eta\to 0}_{xx}(\tilde \omega^{\prime}_{\rm pr})]}{\omega^{2}_{\rm pr} -\omega^{\prime ~ 2}_{\rm pr}} d\omega^{\prime}_{\rm pr}~, \nonumber \\
     &=\dfrac{ e^2 \alpha^2}{h} \dfrac{5 \Gamma[-\frac{5}{4}]^2}{64 \sqrt{2 \pi}\alpha^3}   \dfrac{2 \tilde{\omega}_{\rm pr}}{\pi} \mathcal{P} \int^{\infty}_{0} \dfrac{\sqrt{\tilde{\omega}^{\prime}_{\rm pr}}\Theta\left(\dfrac{\tilde{\omega}^{\prime}_{\rm pr}}{2}-\tilde\mu\right)}{\tilde{\omega}_{\rm pr}^2 -\tilde{\omega}^{\prime~2}_{\rm pr}} d\tilde{\omega}^{\prime}_{\rm pr}~,\nonumber\\
     &=\dfrac{ e^2}{\pi h} \dfrac{5 \sqrt{\tilde{\omega}_{\rm pr}} \Gamma[-\frac{5}{4}]^2}{32  \sqrt{2 \pi}\alpha}  \left[\dfrac{1}{2}\ln{\dfrac{\sqrt{\tilde{\omega}_{\rm pr}}+\sqrt{\tilde{\omega}^{\prime}_{\rm pr}}}{\sqrt{\tilde{\omega}_{\rm pr}}-\sqrt{\tilde{\omega}_{\rm pr}'}}}-\tan^{-1}(\sqrt{\dfrac{\tilde{\omega}^{\prime}_{\rm pr}}{\tilde{\omega}_{\rm pr}}})\right]^{2\tilde\lambda}_{2|\tilde\mu|} 
\end{align}
and, 
\begin{align}
     &{\rm Im}[\sigma^{\eta\to 0}_{yy}( \omega_{\rm pr})]\nonumber\\
     &=\dfrac{2 \omega_{\rm pr}}{\pi} \mathcal{P} \int^{\infty}_{0} \dfrac{{\rm Re}[\sigma^{\eta\to 0}_{yy}(\tilde \omega^{\prime}_{\rm pr})]}{\omega^{2}_{\rm pr} -\omega^{\prime~2}_{\rm pr}} d\omega^{\prime}_{\rm pr}~,\nonumber\\
     &=\dfrac{ e^2 \alpha^2}{h}\dfrac{ \Gamma[\frac{1}{4}]^2}{6 \sqrt{2 \pi}\alpha} \dfrac{2 \tilde{\omega}_{\rm pr}}{\pi} \mathcal{P} \int^{\infty}_{0} \dfrac{\Theta\left(\dfrac{\tilde{\omega}^{\prime}_{\rm pr}}{2}-\tilde\mu\right)}{\sqrt{\tilde{\omega}^{\prime}_{\rm pr}}(\tilde{\omega}_{\rm pr}^2 -\tilde{\omega}_{\rm pr}^{\prime~2})} d\tilde{\omega}^{\prime}_{\rm pr}~,\nonumber\\
     &=\dfrac{ e^2 }{ h\pi }\dfrac{\alpha \Gamma[\frac{1}{4}]^2}{3 \sqrt{2 \pi \tilde{\omega}_{\rm pr}}}\left[\dfrac{1}{2}\ln{\dfrac{\sqrt{\tilde{\omega}_{\rm pr}}+\sqrt{\tilde{\omega}^{\prime}_{\rm pr}}}{\sqrt{\tilde{\omega}_{\rm pr}}-\sqrt{\tilde{\omega}^{\prime}_{\rm pr}}}}+\tan^{-1}(\sqrt{\dfrac{\tilde{\omega}_{\rm pr}^{\prime}}{\tilde{\omega}_{\rm pr}}})\right]^{2\tilde\lambda}_{2|\tilde\mu|}
\end{align}
where we have taken the upper limit of integration as the ultraviolet cutoff defined by $\lambda (= \tilde\lambda t)$.
%
\subsection{Nonlinear regime}
For the $\eta^2$ contribution to the real parts of the optical conductivity, we can proceed as before so that we have, 
\begin{align}
    {\rm Im}&[\sigma^{\eta^2}_{xx}(\omega_{\rm pr}; \omega)]\nonumber\\
    &=\frac{2e^2}{\pi h} \bigg[\dfrac{5 \sqrt{\tilde{\omega}_{\rm pr}} \Gamma[-\frac{5}{4}]^2}{64 \pi \sqrt{2 \pi}\alpha}  \left[\dfrac{1}{2}\ln{\dfrac{\sqrt{\tilde{\omega}_{\rm pr}}+\sqrt{\tilde{\omega}^{\prime}_{\rm pr}}}{\sqrt{\tilde{\omega}_{\rm pr}}-\sqrt{\tilde{\omega}_{\rm pr}'}}}- \right. \nonumber\\
    &\left. \tan^{-1}(\sqrt{\dfrac{\tilde{\omega}^{\prime}_{\rm pr}}{\tilde{\omega}_{\rm pr}}})\right]^{2\tilde\lambda}_{2|\tilde\mu|}  -\frac{2 \alpha \tilde \omega_{pr} \tilde \omega^{2} \tilde{\gamma}_2 \eta^2 \Gamma[\frac{3}{4}]^2 \sin^2{\theta}}{15\sqrt{\pi}} \nonumber \\
    & \left[\frac{4 \tilde{\gamma}_2}{\sqrt{\tilde{\omega}_{pr}^\prime} \tilde{\omega}_{pr}^2 \left(\tilde{\gamma}_2^2+\tilde{\omega}^2\right)} + \frac{2 A}{\tilde{\omega}_{pr}^{5/2}} \left[ \dfrac{1}{2}\ln{\dfrac{\sqrt{\tilde{\omega}_{\rm pr}}+\sqrt{\tilde{\omega}^{\prime}_{\rm pr}}}{\sqrt{\tilde{\omega}_{\rm pr}}-\sqrt{\tilde{\omega}_{\rm pr}'}}}-\right.\right.\nonumber\\
    &\left.\left.\tan^{-1}(\sqrt{\dfrac{\tilde{\omega}^{\prime}_{\rm pr}}{\tilde{\omega}_{\rm pr}}}) \right] + B(\frac{3}{2},\tilde{\omega}) + B(\frac{3}{2},-\tilde{\omega}) \right]^{2\tilde\lambda}_{2|\tilde\mu|}  - \nonumber \\
    & \frac{63 \tilde{\omega}_{pr} \tilde{\omega}^{2} \tilde{\gamma}_2 \eta^2 \Gamma[-\frac{7}{4}]^2 \cos^2{\theta}}{5632  \alpha^2 \sqrt{2\pi}} \left[ \frac{2A}{\tilde{\omega}_{pr}^{3/2}} \left[ \dfrac{1}{2}\ln{\dfrac{\sqrt{\tilde{\omega}_{\rm pr}}+\sqrt{\tilde{\omega}^{\prime}_{\rm pr}}}{\sqrt{\tilde{\omega}_{\rm pr}}-\sqrt{\tilde{\omega}_{\rm pr}'}}} \right. \right. \nonumber \\
    &\left. \left. +\tan^{-1}(\sqrt{\dfrac{\tilde{\omega}^{\prime}_{\rm pr}}{\tilde{\omega}_{\rm pr}}}) \right] -B(\frac{1}{2},\tilde{\omega}) - B(\frac{1}{2},-\tilde{\omega}) \right]^{2\tilde\lambda}_{2|\tilde\mu|}\bigg] ,
\end{align}
\begin{align}
    {\rm Im}&[\sigma^{\eta^2}_{yy}(\omega_{\rm pr}; \omega)] =\frac{2e^2}{\pi h} \bigg[\dfrac{ \Gamma[\alpha \frac{1}{4}]^2}{6 \pi \sqrt{2 \pi \tilde{\omega}_{\rm pr}}} \left[\dfrac{1}{2}\ln{\dfrac{\sqrt{\tilde{\omega}_{\rm pr}}+\sqrt{\tilde{\omega}^{\prime}_{\rm pr}}}{\sqrt{\tilde{\omega}_{\rm pr}}-\sqrt{\tilde{\omega}^{\prime}_{\rm pr}}}} \right. \nonumber\\
    &\left. +\tan^{-1}(\sqrt{\dfrac{\tilde{\omega}_{\rm pr}^{\prime}}{\tilde{\omega}_{\rm pr}}})\right]^{2\tilde\lambda}_{2|\tilde\mu|}  -\frac{15 \alpha^3 \tilde{\omega}_{pr} \tilde{\omega}^{2} \tilde{\gamma}_2 \eta^2 \Gamma[-\frac{3}{4}]^2 \sin^2{\theta}}{224\sqrt{2 \pi}} \nonumber\\
    &  \left[\frac{4 \tilde{\gamma}_2}{3\sqrt{\tilde{\omega}_{pr}^\prime} \tilde{\omega}_{pr}^2 \left(\tilde{\gamma}_2^2+\tilde{\omega}^2\right)} + \frac{2 A}{\tilde{\omega}_{pr}^{7/2}} \left[ \dfrac{1}{2}\ln{\dfrac{\sqrt{\tilde{\omega}_{\rm pr}}+\sqrt{\tilde{\omega}^{\prime}_{\rm pr}}}{\sqrt{\tilde{\omega}_{\rm pr}}-\sqrt{\tilde{\omega}_{\rm pr}'}}} \right. \right.\nonumber \\
    & \left. \left. +\tan^{-1}(\sqrt{\dfrac{\tilde{\omega}^{\prime}_{\rm pr}}{\tilde{\omega}_{\rm pr}}}) \right] - B(\frac{5}{2},\tilde{\omega}) - B(\frac{5}{2},-\tilde{\omega}) \right]^{2\tilde\lambda}_{2|\tilde\mu|}  - \nonumber \\
    & \frac{4 \alpha \tilde{\omega}_{pr} \tilde{\omega}^{2} \tilde{\gamma}_2 \eta^2 \Gamma[\frac{3}{4}]^2 \cos^2{\theta}}{15  \alpha^2 \sqrt{2\pi}} \left[\frac{4 \tilde{\gamma}_2}{\sqrt{\tilde{\omega}_{pr}^\prime} \tilde{\omega}_{pr}^2 \left(\tilde{\gamma}_2^2+\tilde{\omega}^2\right)}  \right.\nonumber \\
    & \left. + \frac{2 A}{\tilde{\omega}_{pr}^{5/2}} \left[ \dfrac{1}{2}\ln{\dfrac{\sqrt{\tilde{\omega}_{\rm pr}}+\sqrt{\tilde{\omega}^{\prime}_{\rm pr}}}{\sqrt{\tilde{\omega}_{\rm pr}}-\sqrt{\tilde{\omega}_{\rm pr}'}}}-\tan^{-1}(\sqrt{\dfrac{\tilde{\omega}^{\prime}_{\rm pr}}{\tilde{\omega}_{\rm pr}}}) \right]  \right. \nonumber \\
    & \left. + B(\frac{3}{2},\tilde{\omega}) + B(\frac{3}{2},-\tilde{\omega}) \right]^{2\tilde\lambda}_{2|\tilde\mu|}\bigg] ~,
\end{align}
and,
\begin{align}
    {\rm Im}&[\sigma^{\eta^2}_{xy}({\omega}_{\rm pr};{\omega})]=-\frac{2 e^2}{\pi h}\frac{2\sqrt{2} \alpha ^3 \tilde{\omega}_{pr} \tilde{\omega}^{2} \tilde{\gamma}_2 \eta^2 \Gamma[\frac{3}{4}]^2 \sin{2\theta}}{15   \sqrt{\pi}} \nonumber\\
    & \left[\frac{4 \tilde{\gamma}_2}{\sqrt{\tilde{\omega}_{pr}^\prime} \tilde{\omega}_{pr}^2 \left(\tilde{\gamma}_2^2+\tilde{\omega}^2\right)} + \frac{2 A}{\tilde{\omega}_{pr}^{5/2}} \left[ \dfrac{1}{2}\ln{\dfrac{\sqrt{\tilde{\omega}_{\rm pr}}+\sqrt{\tilde{\omega}^{\prime}_{\rm pr}}}{\sqrt{\tilde{\omega}_{\rm pr}}-\sqrt{\tilde{\omega}_{\rm pr}'}}} \right. \right.\nonumber \\
    & \left. \left. -\tan^{-1}(\sqrt{\dfrac{\tilde{\omega}^{\prime}_{\rm pr}}{\tilde{\omega}_{\rm pr}}}) \right] + B(\frac{3}{2},\tilde{\omega}) + B(\frac{3}{2},-\tilde{\omega}) \right] ^{2\tilde\lambda}_{2|\tilde\mu|} 
\end{align}

where,
\begin{align}
    &A=\frac{2 \tilde{\gamma}_2 \left(\tilde{\gamma}_{2}^2+\tilde{\omega}^2+\tilde{\omega}_{pr}^2\right)}{ \left(\tilde{\gamma}_{2}^4+2 \tilde{\gamma}_{2}^2 \left(\tilde{\omega}^2+\tilde{\omega}_{pr}^{2}\right)+\left(\tilde{\omega}^2-\tilde{\omega}_{pr}^{2}\right)^2\right)} \\
    &B(x,\tilde{\omega}) = \frac{i \left(\frac{\tan ^{-1}\left(\frac{\sqrt{\tilde{\omega}_{pr}^\prime}}{\sqrt{-\tilde{\omega}-i \tilde{\gamma}_2}}\right)}{(-\tilde{\omega}-i \tilde{\gamma}_2)^{x}}-\frac{\tan ^{-1}\left(\frac{\sqrt{\tilde{\omega}_{pr}^\prime}}{\sqrt{\tilde{\omega}+i \tilde{\gamma}_2}}\right)}{(\tilde{\omega}+i \tilde{\gamma}_2)^{x}}\right)}{\tilde{\omega}_{pr}^2-(\tilde{\omega}+i \tilde{\gamma}_{2})^2}
\end{align}

\subsection{Intra-band optical conductivity}
In the main text we have discussed only the inter-band part of the optical conductivity. Here, for simplicity we provide the results for the intra-band part for $\Delta = 0$ case. This can be easily generalized for finite $\Delta$ cases.
The intra-band conductivity can be given as,
\begin{eqnarray}
\frac{\sigma^{\rm D}_{ij}(\omega_{\rm pr};\omega)}{
    g_s/{(4\pi^2\hbar)}}=- 2i \int \frac{d{\bf k}}{\omega_{\bf k}} \frac{\partial n_{\bf k}}{\partial\omega_{\bf k}}\frac{{\mathcal M}^{cc}_{{\bf k} i}{\mathcal M}^{cc}_{{\bf k}j}}{ \omega_{\rm pr} + i\gamma}~.
\end{eqnarray}   
Now for the linear regime $ n_{\bf k}$ is replace by $ n^{\rm eq}_{\bf k}$ such that for $T=0$, $n^{\rm eq}_{\bf k}= \Theta\left(\mu-\hbar\omega_{\bf{k}}/2\right)-\Theta\left(\mu+\hbar\omega_{\bf{k}}/2\right)$. Then the longitudinal conductivities for intraband at $T=0$ are given by,
\begin{eqnarray}
    \sigma^{{\rm D},\eta\to 0}_{xx}( \omega_{\rm pr})\nonumber\\
    =-2i&&\frac
    {g_s}{(2\pi)^{2}\hbar} \int d{\bf k} \frac{\partial n^{\rm eq}_{\bf k}}{\partial\omega_{\bf k}}\frac{{\mathcal M}^{cc}_{{\bf k} x}{\mathcal M}^{cc}_{{\bf k}x}}{ \omega_{\rm pr} + i\gamma},\nonumber\\
    =2i&&\frac 
    {g_s}{(2\pi)^{2}\hbar} \int d{\bf k} \frac{\hbar}{2}\delta\left(\frac{\hbar\omega_{\bf{k}}}{2}-\mu\right)\frac{{\mathcal M}^{cc}_{{\bf k} x}{\mathcal M}^{cc}_{{\bf k}x}}{ \omega_{\rm pr} + i\gamma},\nonumber\\
    =\frac{e^2}{\pi h} &&\int \frac{\tilde \omega_{\bf k} ~d \tilde \omega_{\bf k}~d\phi}{S}\delta(\frac{\tilde \omega_{\bf{k}}}{2}-\tilde \mu) \frac{(S-X)^{3}}{\tilde \omega^{2}_{\bf{k}} \cos^{6}\phi}\frac{i}{ \tilde \omega_{\rm pr} + \tilde \gamma},\nonumber\\
    =\frac{e^2}{\pi h}&&\frac{15 \tilde\mu^{3/2} \Gamma \left(-\frac{5}{4}\right)^{2}}{32 \sqrt{ \pi } \alpha }\frac{i}{ \tilde \omega_{\rm pr} + \tilde \gamma}~,
\end{eqnarray}
and, 
\begin{eqnarray}
    \sigma^{{\rm D},\eta\to 0}_{yy}(\omega_{\rm pr})\nonumber\\
    =-2i&&\frac
    {g_s}{(2\pi)^{2}\hbar} \int d{\bf k} \frac{\partial n^{\rm eq}_{\bf k}}{\partial\omega_{\bf k}}\frac{{\mathcal M}^{cc}_{{\bf k} y}{\mathcal M}^{cc}_{{\bf k}y}}{ \omega_{\rm pr} + i\gamma},\nonumber\\
    =2i&&\frac
    {g_s}{(2\pi)^{2}\hbar} \int d{\bf k} \frac{\hbar}{2}\delta(\frac{\hbar\omega_{\bf{k}}}{2}-\mu)\frac{{\mathcal M}^{cc}_{{\bf k} y}{\mathcal M}^{cc}_{{\bf k}y}}{\omega_{\rm pr} + i\gamma},\nonumber\\
    =\frac{ie^2}{\pi h} &&\int \frac{\tilde \omega_{\bf k} ~d \tilde \omega_{\bf k}~d\phi_{\bf k}}{S}\delta(\frac{\tilde \omega_{\bf{k}}}{2}-\tilde \mu) \frac{4 \alpha^{4} \sin^{2}\phi_{\bf k} (S-X)}{\tilde \omega^{2}_{\bf{k}} \cos^{4}\phi_{\bf k}(\omega_{\rm pr} + i\gamma)},\nonumber\\
    =\frac{e^2}{\pi h}&&\frac{3 \alpha \Gamma \left(-\frac{3}{4}\right)^2 \sqrt{\tilde\mu}}{8 \sqrt{ \pi }}\frac{i}{ \tilde \omega_{\rm pr} + \tilde \gamma}~.
\end{eqnarray}
For the nonlinear case, we can proceed as we did for the inter-band optical conductivities. However, it yields a very lengthy expression due to the factor $\partial_{\omega_{\bf k}}n_{\bf k}$.
%
%
%
\bibliography{Ref}
\end{document}